\documentclass{aa}  
\usepackage{natbib}
\usepackage{graphicx}
\usepackage{txfonts}
\usepackage[normalem]{ulem}
\usepackage{xcolor}
\usepackage{caption}
\usepackage{subcaption} 
\usepackage{placeins} 

\definecolor{links}{rgb}{0.11, 0.67, 0.84}
\usepackage[colorlinks=true,allcolors=links]{hyperref}

\definecolor{NA}{rgb}{0.64,0.20,0.64} 

\definecolor{ML}{rgb}{0.09,0.11,0.50} 

\definecolor{DG}{rgb}{0.0, 0.8, 0.6}

\begin{document} 
  \title{Relative distribution of dark matter, gas, and stars around cosmic filaments in the IllustrisTNG simulation}
  
  \titlerunning{Matter in cosmic filaments}
  
  \author{Daniela Gal\'arraga-Espinosa\inst{1,2}
    \and Mathieu Langer\inst{1}
    \and Nabila Aghanim\inst{1}
        }
        

    \institute{Universit\'e Paris-Saclay, CNRS, Institut d'astrophysique spatiale, 91405, Orsay, France
    \and 
    Max-Planck Institute for Astrophysics, Karl-Schwarzschild-Str.~1, D-85741 Garching, Germany\\
     \email{danigaes@mpa-garching.mpg.de}}

  \date{Received XXX; accepted YYY}

  \abstract{
   {{We present a comprehensive study of the distribution of matter around different populations of large-scale cosmic filaments, using the IllustrisTNG simulation at $z=0$. We computed the dark matter (DM), gas, and stellar radial density profiles of filaments, and we characterise the distribution of the baryon fraction in these structures. We find that baryons exactly follow the underlying DM distribution only down to $r \sim 7$ Mpc to the filament spines. At shorter distances ($r < 7$ Mpc), the baryon fraction profile of filaments departs from the cosmic value $\Omega_\mathrm{b} / \Omega_\mathrm{m}$. While in the $r \sim 0.7-7$ Mpc radial domain this departure is due to the radial accretion of the warm-hot intergalactic medium (WHIM) towards the filament cores (creating an excess of baryons with respect to the cosmic fraction), the cores of filaments ($r < 0.7$ Mpc) show a clear baryon depletion instead. The analysis of the efficiency of active galactic nuclei (AGN) feedback events in filaments reveals that they are potentially powerful enough to eject gas outside of the gravitational potential wells of filaments.
   We show that the large-scale environment (i.e. denser versus less dense, hotter versus colder regions) has a non-negligible effect on the absolute values of the DM, gas, and stellar densities around filaments. Nevertheless, the relative
   distribution of baryons with respect to the underlying DM density field is found to be independent of the filament population. Finally, we provide scaling relations between the gas density, temperature, and pressure for the different populations of cosmic filaments. We compare these relations to those pertaining to clusters of galaxies, and find that these cosmic structures occupy separate regions of the density-temperature and density-pressure planes.}
    }}


\maketitle
\newpage


\section{Introduction}

On the largest scales of the Universe, matter is organised in nodes, filaments, walls, and voids, creating a gigantic network called the cosmic web \citep{Bond1996}. This network is mostly made of dark matter (DM) which, from the initial fluctuations of the primordial density field, collapsed under the action of gravity and formed the cosmic filamentary pattern \citep{Zeldovich1970} that we observe today. In the cosmic web, matter flows from the less dense to the denser environments, going from voids to walls, from walls to filaments, and from filaments to nodes, where it accumulates.

The study of the cosmic skeleton, which is gravitationally shaped by DM, has been made possible thanks to the development of \textit{N}-body numerical simulations \citep[e.g.][]{Springel2005, BoylanKolchin2009_MilleniumII, Klypin2011_Bolshoi_sim, Prada2012_MultiDark_sim,Potter2017_EuclidNbody_sim}, and the development of web-finder methods \citep[for e.g.][]{Libeskind2018}. For example, the seminal studies of \cite{AragonCalvo2010} and \citet{Cautin2014} have shown that most of the cosmic volume at $z=0$ is occupied by voids, and, despite nodes being the denser cosmic structures, $\sim 50 \%$ of today's cosmic mass resides in filaments.
Also, among other results, these papers have shown overlapping ranges for the DM density distributions in the different cosmic environments. For example, encompassing a broad range of almost five orders of magnitude, the density distribution in filaments partially overlaps with that in nodes towards the higher density values, and with that in voids towards the lower ones, hinting at the diversity of the filamentary structures in the cosmic web.

Baryons account for $\sim 15.7 \%$ of the total budget of matter \citep{Planck2015Cosmo}. Ruled by gravity, they fall into the gravitational potential wells created by DM, and they are distributed around the cosmic skeleton. Despite baryons being the only matter component that is directly observable, their signal in cosmic filaments is only difficultly detected in actual observational data. This is due to the small matter densities and complex morphologies of these structures with respect to the denser, more luminous (and thoroughly studied) clusters of galaxies.
The detection of baryons around filaments thus represents a challenge which demands different observational approaches and techniques.
For example, by observing individual pairs of clusters, \citet{TittleyHenriksen2001}, \citet{Werner2008}, \citet{PlanckCollaboration2013_filaments}, \citet{Sugawara2017}, \citet{Akamatsu2017}, \citet{Alvarez2018}, \citet{Bonjean2018}, \citet{Connor2018}, \citet{Connor2019}, \citet{Govoni2019}, \citet{Hincks2022_A399A401}, and \citet{Biffi2021_eROSITAfils} (among others) have detected filaments of scales of a few megaparsecs, acting as bridges of matter between the cluster pairs. Using statistical approaches, some of the properties of these bridges (such as their density, temperature, or magnetic field strength) have been constrained thanks to stacking techniques \citep[e.g.][]{Tanimura2019, DeGraaff2019, Vernstrom2021_LOFAR}. On scales of tens of megaparsecs, longer filaments have been detected by applying different filament finders \citep[e.g.][]{Disperse_paper1, Sousbie2011b, Cautun2013nexus, Tempel2016bisous, Bonnaire2020Trex} to galaxy surveys. This has enabled the study of the properties of galaxies \citep[e.g.][]{Alpaslan2016_GAMAgalaxies, Malavasi2017, Chen2017_fil_gal, Laigle2018, Kraljic2018, Sarron2019, Malavasi2020_sdss, Rost2020, Bonjean2020filaments, Welker2020_sami}, and gas in filaments \citep{Tanimura2020_byopic, Tanimura2020_Rosat}. More precisely, the first constrains on the gas density and temperature of large-scale filaments have been recently obtained by \cite{Tanimura2020_byopic} and \cite{Tanimura2020_Rosat} thanks to the combined analysis of Sunyaev-Zel’dovich (SZ) and lensing signals, and of stacked X-ray data. Also, filaments connecting clusters of galaxies have been observed by focusing on the outskirts of clusters, at small scales in, for example, \cite{Eckert2015_nature}, but also on larger scales \citep[e.g.][]{Malavasi2020_coma}. 

The previous studies concern the local Universe, but several efforts have also been made in order to detect filaments at higher redshifts. For example, thanks to the careful analysis of the Lyman-alpha emission, small-scale filaments connecting galaxies at $z\sim 3-4$ have been observed by \cite{Gallego2018_Lya_fils}, and a detection of larger-scale structures has been reported by \cite{Kikuta2019_Lya_fils}. Other observations at these high redshifts have also been possible thanks to Lyman-alpha forest tomography \citep[e.g.][]{Lee2014_Lya_fils, Japelj2019_Lya_fils}, and predictions concerning the potential detection of large-scale filaments using the 21 cm signal of neutral hydrogen have been stated by \citet{Kooistra2019_21cm_fils}. All these studies have revealed both the multifaceted and multi-scale aspect of the filaments of the cosmic web, and the need to further characterise their properties in order to better understand these structures.\\

Thanks to the development of new numerical methods and of baryonic models calibrated on observations, the increasingly accurate hydro-dynamical simulations \citep[e.g.][]{Dubois2014, Hirschmann2014MAGNETICUM, Dolag2015MAGNETICUM, Schaye2015_EAGLEsimu, McCarthy2017_BAHAMAS, Nelson2019_TNGdata_release, Dave2019_SIMBAsim} have recently enabled the `theoretical' study of cosmic filaments and of the properties of matter within them. We mention, for example, the work of \cite{GhellerVazza2015, GhellerVazza2016, GhellerVazza2019_surveyTandNTprops_fils, Rost2020, GalarragaEspinosa2020} aiming at characterising these structures, and the particular interest on their gas content presented in \cite{Martizzi2019a, Tuominen2021, GalarragaEspinosa2021}. Moreover, the analysis of different properties of galaxies in filaments has been performed by e.g.~\cite{Kraljic2018, Kraljic2019, Kraljic2020spin, GaneshaiahVeena2019_CosmicBalletII, Song2021}, and \cite{Gouin2020, Kraljic2020, Gouin2021, Rost2021, Gouin2022} have studied how these filamentary structures connect to clusters in hydro-dynamical simulations.\\

In this work, we use the outputs of the IllustrisTNG simulation \citep{Nelson2019_TNGdata_release} and, by the means of radial density profiles, we perform the first comprehensive analysis of the relative distribution of all the matter components (i.e. DM, gas, and stars) around the large-scale ($\sim \, \mathrm{Mpc}$) cosmic filaments.
Based on the studies of the galaxy distribution and gas properties around these structures \citep{GalarragaEspinosa2020, GalarragaEspinosa2021}, filaments are separated in two populations: the short, tracers of denser and hotter regions, and the long filaments, tracers of less dense and colder environments. A special attention is put on the study of the relative behaviours of the DM, gas, and stellar profiles, as well as on their shapes as a function of the filament population. We also investigate the evolution of the baryon fraction in filaments as a function to the distance to the filament spines and we present scaling relations between the gas density of filaments and some observables, namely gas temperature and pressure.

The IllustrisTNG simulation is briefly introduced in Sect.~\ref{Sect:Data}, along with the filament catalogue used in this work. Section~\ref{Sect:Densities} shows the density and over-density profiles, and the study of the baryon fraction in filaments is presented in Sect.~\ref{Sect:BaryonFrac}. Finally, the resulting scaling relations are provided in Sect.~\ref{Sect:Scalings}, and we summarise our main conclusions in Sect.~\ref{Sect:Conclusions}.

\section{\label{Sect:Data}Data}

\subsection{\label{SubSect:IllustrisTNG}IllustrisTNG simulation}

We use the outputs of the gravo-magnetohydrodynamical simulation IllustrisTNG\footnote{\url{https://www.tng-project.org}} \citep{Nelson2019_TNGdata_release}, which follows the coupled evolution of DM, gas, stars, and black holes from redshift $z=127$ to $z=0$. This simulation is run with the moving-mesh code Arepo \citep{Arepo}, and the values of the cosmological parameters are those of \cite{Planck2015Cosmo}:  $\Omega_{\Lambda,0} = 0.6911$, $\Omega_{m,0}=0.3089$, $\Omega_{b,0}=0.0486$, $\sigma_{8}=0.8159$, $n_s=0.9667$ and $h=0.6774$.
We study the TNG300-1 simulation box, which is the best resolution run ($m_{\mathrm{DM}} = 4.0 \times 10^{7} \mathrm{M_{\odot}}/h$) of the largest volume of the IllustrisTNG simulation.
This box, which consists of a cube of about 302 Mpc side length, allows both the statistically significant detection of long cosmic filaments and the study of processes down to small scales.
In the following, all the results are derived from the snapshot at redshift $z=0$.

The gaseous component in the IllustrisTNG simulation is implemented by Voronoi cells that evolve in time, refined and de-refined according to a mass target of $7.6 \times 10^6 \mathrm{M}_\odot/h$ \citep{Arepo, Nelson2019_TNGdata_release, Weinberger2020arepo, Pillepich2018TNGmodel}, and the baryonic processes driving gas dynamics (like star formation, stellar evolution, chemical enrichment, gas cooling, black hole formation, growth, and feedback, etc.) are described by the `TNG model' \citep{Pillepich2018TNGmodel, Nelson2019_TNGdata_release}, implemented in a subgrid manner, whose description and details can be found in \cite{Pillepich2018TNGmodel}. We note that this model was specifically calibrated on observational data to match the observed galaxy properties and statistics \citep{Nelson2019_TNGdata_release}.

\subsection{\label{SubSect:Analysed_data_sets}Analysed data sets}

The full snapshot of the TNG300-1 box at $z=0$ contains a very large number of DM particles and gas cells ($\sim 14 \times 10^{9}$), and $\sim 7 \times 10^{8}$ star particles. Running the analysis on this extremely large amount of information is computationally very expensive and time consuming, so in order to compute the density profiles in reasonable computing time, we sample the DM, star particles, and the gas cells by randomly selecting one out of 1000 (and we rescale the resulting densities by this factor). As shown in Appendix~\ref{Appendix:Subsampling}, this sub-sampling factor turns out to be a good compromise between computational speed and precision, and it does not bias the density estimates with respect to the full particle distribution. Moreover, performing the analysis on different random extractions gives stable results, so this sub-sampling does not undermine the statistical conclusions presented in this work.

\subsection{\label{SubSect:filament_catalogue}The filament catalogue}

    \begin{figure*}
    \centering
   \includegraphics[width=1\textwidth]{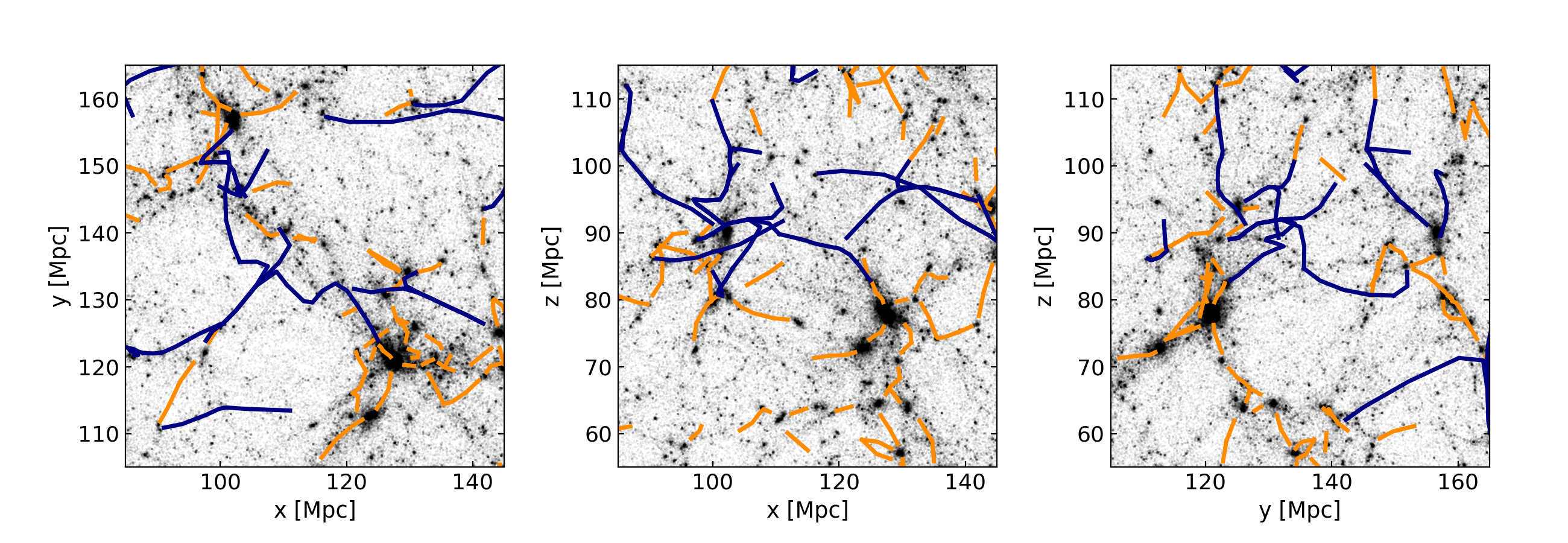}
   \caption{Box of 60 Mpc side length, projections onto the \textit{xy},\textit{xz}, \textit{yz} planes. Orange and blue lines represent the spines of short and long filaments, respectively. The black points show the underlying DM distribution. Short filaments are at the vicinity of nodes, whereas the long population traces less dense regions.}
    \label{Fig:visu_fils}
    \end{figure*}
    
We detect the filaments in the TNG300-1 simulation box 
by applying the DisPerSE code \citep{Disperse_paper1, Sousbie2011b} to the galaxy distribution at $z = 0$. DisPerSE identifies filaments from the topology of the density field, traced here by galaxies, using Discrete Morse theory. All the details and technicalities of the extraction of the skeleton used in this work can be found in \cite{GalarragaEspinosa2020}, where the catalogue of filaments was first constructed.
It is important to mention that, following \cite{GalarragaEspinosa2020, GalarragaEspinosa2021}, a filament is defined as the set of straight segments connecting a point of maximum density (hereafter CPmax), to a saddle of the density field \cite[see Fig.~2 of][]{GalarragaEspinosa2020}.
Within this definition, the filaments of our catalogue have a maximum length of $L_f = 65.6$ Mpc, a minimum of and $L_f = 0.4$ Mpc, and their mean and median lengths are respectively $10.9$ and $8.8$ Mpc. Moreover, since this study focuses on the main stem of filaments and not on their connection with nodes, we remove from our catalogue all the filament segments lying closer than $3 \times R_{200}$ to the centre of the CPmax points, the topological nodes found by DisPerSE.

The previous studies of \cite{GalarragaEspinosa2020, GalarragaEspinosa2021} have shown that filaments can be classified into two populations, that are the short ($L_f < 9$ Mpc) and the long ($L_f \geq 20$ Mpc) filaments. From the total of 5550 filaments of the catalogue, 2846 are short and 632 are long. Figure \ref{Fig:visu_fils} shows the \textit{xy}, \textit{xz} and \textit{yz} projections of the positions of short and long filaments (respectively in orange and blue) in a sub-box of the TNG300-1 simulation. 
In this figure we can clearly see that short filaments are located at the vicinity of clusters of galaxies (densest regions), while long filaments, living in less dense regions, connect vaster regions. These different environments reflect the findings of \cite{GalarragaEspinosa2020, GalarragaEspinosa2021}, where we have shown that short filaments are denser in gas and galaxies, and that they have on average higher temperature and pressure values than the long population. In the following, we present a comprehensive study of the distribution of matter around the two populations of filaments.

\section{\label{Sect:Densities}Densities of DM, gas, and stars around filaments}

In this Section, we investigate how the different matter components (DM, gas, and stars) are distributed around cosmic filaments. We first compute the radial densities of these three matter components around short and long filaments (Sect.~\ref{SubSect:Densities_absolute}), and then we focus on the relative radial behaviour of DM, gas, and stars by studying their over-density profiles (Sect.~\ref{SubSect:Shapes}).

\subsection{\label{SubSect:Densities_method}Method}

We compute the density profiles of the matter component $i$ ($i=$ DM, gas, or stars) around filaments, by averaging the profiles of individual filaments so that, for a number $N_\mathrm{fil}$ of filaments, the average density of $i$ at a distance $r$ from the filament spines is given by:
\begin{equation}\label{Eq:RHOmean}
    \rho_i(r) \equiv \frac{1}{N_\mathrm{fil}}{ \sum_{f=1}^{N_\mathrm{fil}} \rho_i^{\mathrm{fil},f}(r) }.
\end{equation}
Here $\rho^{\mathrm{fil}, f}(r)$ denote the individual profiles, which are obtained as follows.
For a given filament $f$, the density of the matter component $i$ is computed by summing its mass enclosed in hollow cylinders around all the $N_\mathrm{seg}$ segments belonging to the filament. Then, this mass is divided by the corresponding volume, as shown by:
\begin{equation}\label{Eq:RHOfils}
    \rho_{i}^{\mathrm{fil}, f}(r_k) = \frac{\displaystyle
    \sum_{s = 1}^{N_\mathrm{seg}}
   M_i^s(r_{k}, r_{k-1})
    }{
    \pi (r^2_{k} - r^2_{k-1}) \,
    \displaystyle \sum_{s = 1}^{N_\mathrm{seg}} l^s,
    }
\end{equation}
where, for the filament segment $s$, the quantity $M_i^s(r_{k}, r_{k-1})$ is the mass enclosed within the cylindrical shell of outer radius $r_{k}$ and of thickness $r_{k} - r_{k-1}$, centred on the segment axis, and $\sum_{s = 1}^{N_\mathrm{seg}} l^s$ corresponds to the total length of the filament.
Therefore, the total density of matter at a distance $r_k$ from the filament spine can be simply computed by summing the individual contributions of DM, gas, and stars, such that $\rho_\mathrm{TOT}(r_{k}) = \rho_\mathrm{DM}(r_{k}) + \rho_\mathrm{gas}(r_{k}) + \rho_\mathrm{*}(r_{k})$.

Finally, in what follows, all the error bars are computed by using the bootstrap method over the number of filaments. These errors thus represent the statistical variance of the densities around our limited sample of filaments.\\

\subsection{\label{SubSect:Densities_absolute}Radial densities}

    \begin{figure}
    \centering
   \includegraphics[width=0.5\textwidth]{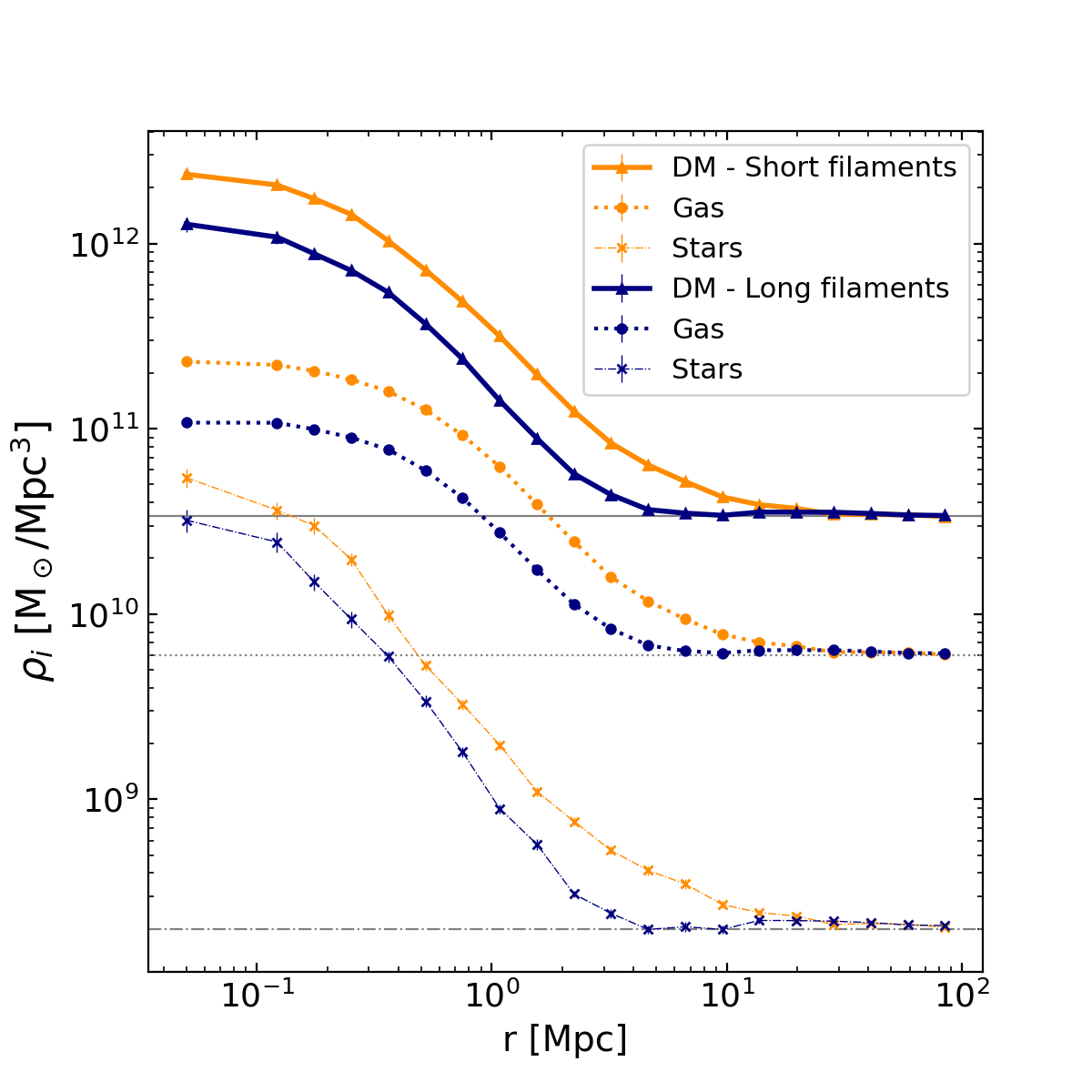}
   \caption{Radial densities of DM (solid lines with triangles), gas (dotted lines with circles) and stars (dash-dotted lines with crosses) around filaments. Results for the short and long populations are presented respectively on orange and blue. The horizontal grey lines correspond to the densities of the matter component $i$ in the full simulation box, $\rho_i^\mathrm{box}$.}
    \label{Fig:density_profiles}
    \end{figure}

The density profiles of DM, gas, and stars around short and long filaments, from their spine up to large radial distances of $\sim 100$ Mpc are presented in Fig.~\ref{Fig:density_profiles}.
First of all, we note that, at large radial scales, all the curves of Fig.~\ref{Fig:density_profiles} reach their respective background density (grey horizontal lines) computed as $\rho^\mathrm{box}_i = M^\mathrm{TOT}_i / V^\mathrm{TOT}_\mathrm{box}$. When approaching the cores of filaments, the densities of these three matter components increase, as expected, and peak at the innermost region of the filaments.

We see a clear difference between the filament populations. For example, in the $r<1$ Mpc region, short filaments are on average $1.95$, $2.10$, and $1.76$ denser than long filaments in DM, gas, and stars, respectively. These results correlate well with the findings of \cite{GalarragaEspinosa2020, GalarragaEspinosa2021}. At $z=0$, short filaments trace denser regions of the cosmic web (like the vicinity of clusters of galaxies), while long filaments, the building blocks of the cosmic skeleton, live in less dense environments.

\subsection{\label{SubSect:Shapes}Over-densities}

    \begin{figure*}
    \centering
    \subfloat[]{%
    \includegraphics[width=1\textwidth]{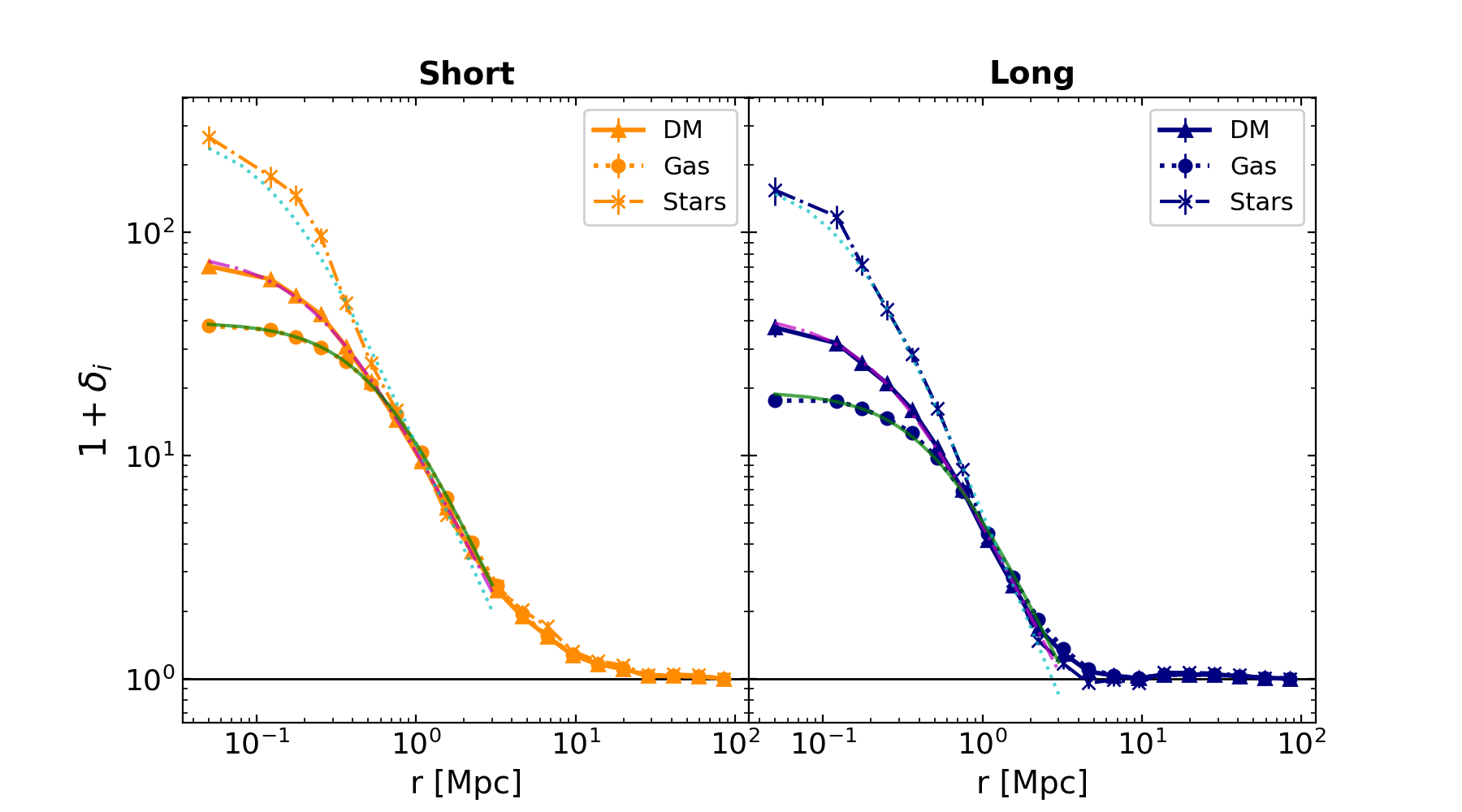}%
    \label{Fig:density_FILS_RESCALED}%
    }\qquad
    \subfloat[]{%
    \includegraphics[width=1\textwidth]{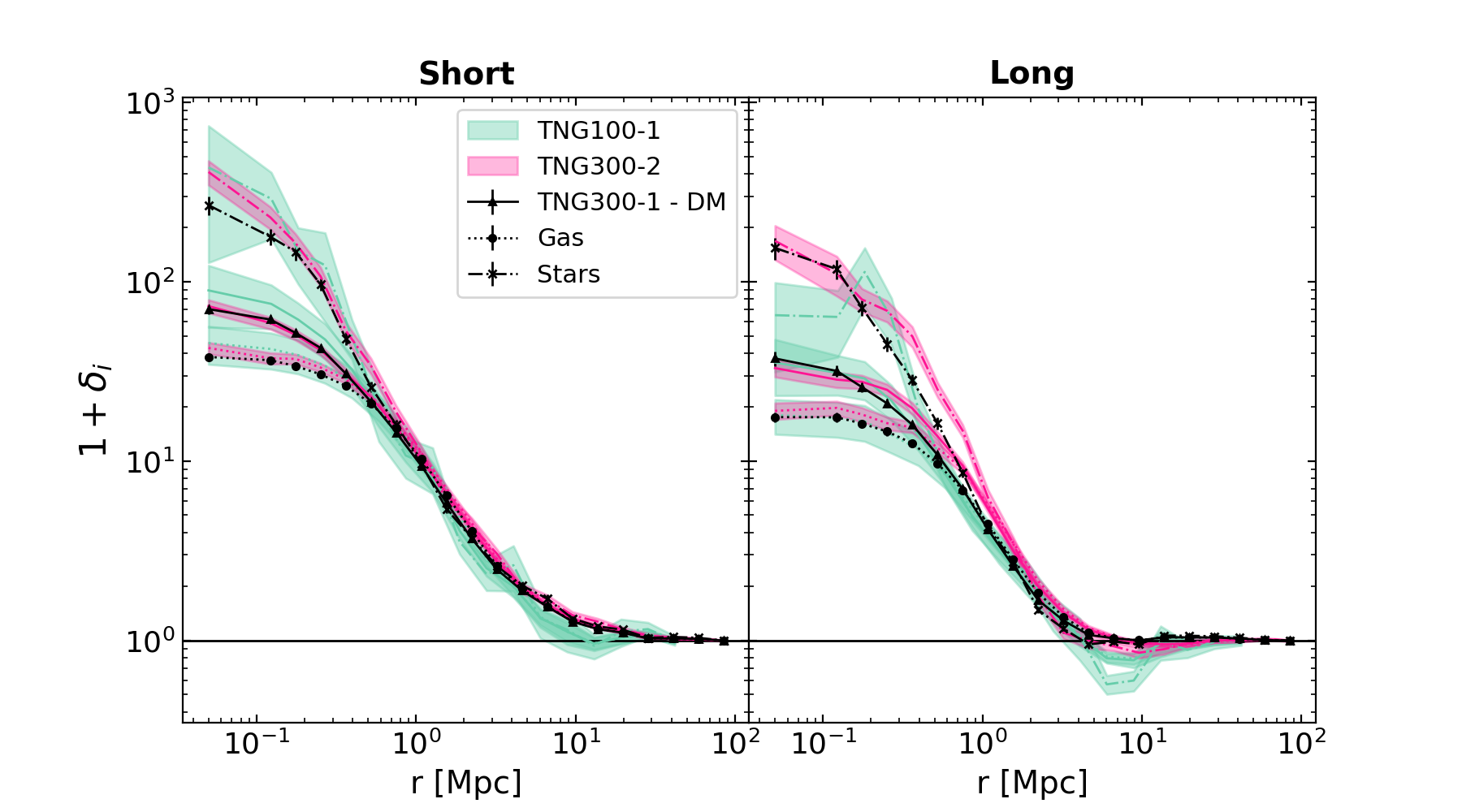}%
    \label{Fig:OVrpo_3sims}%
    }
    \caption{\textit{Top panel:} Radial profiles of the mean over-densities (Eq.~\ref{Eq:OV}) of DM (solid lines with triangles), gas (dotted lines with circles), and stars (dash-dotted lines with crosses) for short and long filaments, respectively in orange and blue. The pink, green, and cyan lines denote respectively the fitting results of the DM, gas, and stellar profiles to the generalised $\beta$-model of Eq.~\ref{Eq:Beta} in the $r < 3$ Mpc range. \textit{Bottom panel:} Impact of the resolution of the simulation. The black curves correspond to the profiles obtained with the reference TNG300-1 simulation (same profiles as in Fig.~\ref{Fig:density_FILS_RESCALED}). Pink and green curves present respectively the results from the TNG300-2 (low resolution) and TNG100-1 (high resolution) boxes.}
    \end{figure*}
    
We now focus on the relative distribution of DM, gas, and stars around filaments. We compute over-density profiles of these three matter components by dividing the densities of Fig.~\ref{Fig:density_profiles} by their respective background value:
\begin{equation}\label{Eq:OV}
    1+\delta_i (r) = \frac{\rho_i(r)}{\rho_i^\mathrm{box}}.
\end{equation}
The over-density profiles are presented in Fig.~\ref{Fig:density_FILS_RESCALED}. Qualitatively, the overlap between the three $1 + \delta_i$ profiles is striking at large distances form the cores of filaments. This shows that, at these distances, gas and stars closely follow the DM distribution and fall, unperturbed, into the DM gravitational potential wells.
Nevertheless, when approaching the cores of filaments, gas and stars clearly depart from the DM over-density. This departure happens at $r \sim 0.7$ Mpc in both short and long filaments. In Sect.~\ref{Sect:BaryonFrac} we see that this radial scale marks a well defined regime in the baryon fraction profiles of filaments. Notice that this scale is smaller than the typical filament thickness at the outskirts of clusters (which are excluded in this work) of \cite{Rost2021}, showing that the filamentary structures connected to clusters of galaxies are different from the main stems of cosmic filaments, as expected for structures in those extremely dense environments of cluster outskirts. Moreover, it is interesting to see that the over-density of gas we find here at the cores of long filaments is compatible with the analysis of SDSS filaments of \cite{Tanimura2020_byopic}, and also in good agreement with the $\delta = 30 \pm 15$ value found by \cite{Tanimura2020_Rosat} using X-ray data. \\

We check the robustness of these results against the resolution of the simulation by performing the same analysis in the TNG300-2 and TNG100-1 simulation boxes, whose resolutions are respectively $\sim 8$ times lower and higher than the reference TNG300-1 . The resulting over-density profiles are shown in Fig.~\ref{Fig:OVrpo_3sims} in black, pink, and green respectively for the TNG300-1, TNG300-2, and TNG100-1 simulations. We note that the black curves are the same as those of Fig.~\ref{Fig:density_FILS_RESCALED}. Apart from the (expected) slight enlargement of the filament cores due to the DisPerSE uncertainty in finding the position of the filament spines in low resolution simulations \citep[see][]{GalarragaEspinosa2020}, the radial profiles of the TNG300-2 filaments are essentially the same as those of the reference simulation, given that their $1 + \delta_i$  values are remarkably close.
The results from TNG300-1 are also compatible with those from the (higher resolution) TNG100-1 box. Nevertheless, the profiles of the latter have very large uncertainties, which is due to the very reduced number of cosmic filaments detected in this small $(110.7 \, \mathrm{Mpc})^3$ box (302 filaments in total, versus 5550 in TNG300-1). As a consequence, from Fig.~\ref{Fig:OVrpo_3sims} it is not possible to detect any potential resolution effects causing density deviations at the filament cores and having amplitudes smaller than the uncertainties of the TNG100-1 profiles.

Still, according to the studies of \cite{Ramsoy2021}, a spatial resolution of at least $l_\mathrm{res} = 2.4$ kpc is needed to have a robust measurement of the densities of cores of (small-scale) filamentary structures of radii $R_\mathrm{res} \sim 8$ kpc. The mean inter-particle distance of the TNG300-1 simulation is $l_\mathrm{TNG} = (V_\mathrm{box}/N_p)^{1/3} = 0.123$ Mpc, so by adopting the $l_\mathrm{res}$ resolution limit, we obtain that TNG300-1 is able to resolve filament of a minimum radius of:
\begin{equation}\label{Eq:Rres}
    R_\mathrm{min} = R_\mathrm{res} \times (l_\mathrm{TNG} / l_\mathrm{res}) = 0.41 \, \mathrm{Mpc},
\end{equation}
which is smaller than the typical $\sim  1$ Mpc radial scales of the cosmic filaments analysed in this work. We therefore deduce that the TNG300-1 simulation is a good compromise between statistics and resolution, providing stable results at the studied scales.

Of course, further studies in higher-resolution boxes with large enough simulated volumes would not only allow for a more detailed analysis of the very inner cores of cosmic filaments, but also to perform a global (multi-scale) study of the filamentary structures of the cosmic web. The latter range from the large (Mpc) scales of cosmic filaments, which are the focus of this work, down to the small (kpc) scales of galactic tendrils, which might be the main providers of gas supplies to the galaxies \citep[e.g.][]{Ramsoy2021}.

\subsection{\label{SubSect:Fit_OV}Modelling the mean DM, gas, and stellar over-density profiles}

\begin{table*}
\caption{Fitting results of the over-density profiles of short filaments (left panel of Fig.~\ref{Fig:density_FILS_RESCALED}) to the $\beta$-model of Eq.~\ref{Eq:Beta}.}
\label{Table:BetaS}     
\centering  
\begin{tabular}{ c | c c c c | c }
 \hline\hline  
     & $(1 + \delta_i^0)$ & $r_0$ [Mpc] & $\alpha$ & $\beta$ & $\chi^2_\nu$ \\ \hline
    Dark matter & $81.66 \pm 5.21$ & $0.25 \pm 0.02$ & $1.41 \pm 0.18$ & $1.00 \pm 0.17$ & $0.39$\\
    Gas & $39.58 \pm 0.89$ & $0.56 \pm 0.05$ & $1.57 \pm 0.12$ & $1.00 \pm 0.12$ & $0.22$ \\ 
    Stars & $287.43 \pm 321.07$ & $0.13 \pm 0.05$ & $1.59 \pm 1.43$ & $1.00 \pm 0.82$ & $4.87$ \\
    \hline
 \end{tabular}
\end{table*}

\begin{table*}
\caption{Fitting results of the over-density profiles of short filaments (right panel of Fig.~\ref{Fig:density_FILS_RESCALED}) to the $\beta$-model of Eq.~\ref{Eq:Beta}.}
\label{Table:BetaL}     
\centering  
\begin{tabular}{ c | c c c c | c }
 \hline\hline  
     & $(1 + \delta_i^0)$ & $r_0$ [Mpc] & $\alpha$ & $\beta$ & $\chi^2_\nu$ \\ \hline
    Dark matter & $43.08 \pm 5.63$ & $0.24 \pm 0.03$ & $1.45 \pm 0.34$ & $1.00 \pm 0.30$ & $0.93$\\
    Gas & $19.34 \pm 1.28$ & $0.51 \pm 0.11$ & $1.53 \pm 0.32$ & $1.00 \pm 0.17$ & $0.88$\\ 
    Stars & $174.40 \pm 17.92$ & $0.14 \pm 0.03$ & $1.73 \pm 0.80$ & $1.00 \pm 0.43$ & $1.75$\\
    \hline
 \end{tabular}
\end{table*}

Let us now analyse the shapes of the profiles shown in Fig.~\ref{Fig:density_FILS_RESCALED}. First, inspired by the findings in small-scale ($\sim$ kpc) filaments connecting galaxies \citep{Ramsoy2021}, we fitted the inner part of the density profiles to a Plummer model. This model, described in Eq.~\ref{Eq:APP_Plummer}, is the theoretical analytic description of an isolated, self-gravitating isothermal cylinder of infinite length \citep{Stodolkiewicz1963_fil_theory, Ostriker1964_fil_theory}. Our results, presented in Appendix~\ref{Appendix:Plummer}, show that the only matter component whose profile is relatively well described by the Plummer model is gas, hinting that gas around the (large-scale) cosmic filaments is in hydrostatic equilibrium, in agreement with the predictions of isothermal cores \citep{KlarMucket2012, GhellerVazza2019_surveyTandNTprops_fils, Tuominen2021, GalarragaEspinosa2021}. Appendix~\ref{Appendix:Plummer} also shows that the mean density of the DM component deviates from the Plummer model, which might hint that filaments are not yet completely self-gravitating structures (because they might still be accreting matter in the radial direction). Nevertheless, given that the filaments in this work have been detected in the galaxy distribution (in order to mimic observations), we abstain from drawing any physical conclusions from this comparison.
Indeed, the spines of filaments extracted from the galaxy distribution might be slightly biased with respect to the density of the underlying DM skeleton. Despite the fact that this bias has been shown to be relatively negligible \citep[][]{Laigle2018}, it could nevertheless be responsible of slightly blurring the profiles at their innermost part, and thus of a possible `non-physical' departure from the Plummer model. Further analyses are thus required in order to conclude on the self-gravitating nature of large-scale cosmic filaments, but we leave this study to a future work. In the remaining of this Section, we model the filament over-density profiles with a more empirical description.\\


We push forward the modelling of the inner part ($r<3$ Mpc) of the over-density profiles of Fig.~\ref{Fig:density_FILS_RESCALED} by releasing the constraints on the slopes of the Plummer profile. This yields the generic $\beta$-model \citep{Cavaliere1976, Arnaud2009, Ettori2013} parameterised by 
\begin{equation}\label{Eq:Beta}
    1 + \delta_i(r) = \frac{ 1 + \delta_i^0 }{ \left(1 + \left(\frac{r}{r_0}\right)^\alpha \right)^\beta },
\end{equation}
and whose resulting fit-curves are shown in pink, green, and cyan, respectively for DM, gas, and stars, in Fig.~\ref{Fig:density_FILS_RESCALED}.
The resulting parameter values for short and long filaments are reported respectively in Tables~\ref{Table:BetaS} and \ref{Table:BetaL}. We see that, for a given matter component, the only parameter that significantly varies from one filament population to the other is the over-density at the core, $1 + \delta^0_i$. Indeed, in addition to the intrinsic differences between matter components, the $1 + \delta^0_i$ values are always larger in short filaments than in long filaments, reflecting the results of Sect.~\ref{SubSect:Densities_absolute}.

Interestingly, the other parameters of this model, namely the radial scale $r_0$ and the slopes $\alpha$ and $\beta$, do not exhibit any major variations between filament populations: their values are rather set by the matter component. 
For example, for both short and long filaments, we find that the $r_0$ scale (marking a transition between the slopes) is smallest in the profiles of stars ($r_0 \sim 0.1$ Mpc), larger in those of DM ($r_0 \sim 0.2$ Mpc), and largest in those of gas ($r_0 \sim 0.5$ Mpc).

Moreover, it is interesting to see that the $\alpha$ parameter, characterising the slopes of the profiles at distances $r>r_0$, is always smaller than $\alpha = 2$. This differs from the trends of DM and gas found in intracluster bridges and in those filaments, sometimes called `spider-legs', that are found at the very outskirts of massive clusters \citep[e.g.][]{Colberg2005, Dolag2006_gas_fils, Rost2021}. These filamentary structures do not belong to the main stem of cosmic filaments, and we recall that they are excluded from our analysis.

At distances $r<r_0$, all the matter components around short and long filaments show an inner slope  $\beta = 1$. This is in line with the findings of \cite{Zhu2021_filaments_evo_z}. However, we note that this regime is close to the resolution scale of the present work. A study at higher resolution is needed in order to penetrate deeper into the filament cores, therefore allowing a better modelling of these regions thanks to a larger number of data points at the inner part of the density profiles.\\

Finally, let us briefly comment on the outer part of the over-density profiles of Fig.~\ref{Fig:density_FILS_RESCALED}, i.e. on the flattening of the profiles at large distances from the spines. This flattening, resulting from the decrease of density at large distances, describes the connection of the filaments with their environment. It happens to be different for the two extreme filament populations. While in long filaments the over-densities reach values of $1+\delta = 1$ at distances as short as $r \sim 4$ Mpc from the spines, the decrease of over-densities is slower in short filaments, where the background value is reached only at $r\sim 12$ Mpc. This slower decrease just reflects the contribution of the denser environment in which the short population is embedded. In order to better understand how cosmic filaments are connected to their environment, it might be interesting to disentangle between the different possible contributions at the outer part of the density profiles of filaments (e.g. walls, other filaments, etc). Nevertheless such study lies outside of the scope of the present work.\\

This section has shown some similarities between the slopes and radial scales of the DM, gas, and stellar density profiles around short and long filaments. This might be a hint that matter is subject to similar dynamics in filaments, despite the different large-scale environments of these two extreme populations (i.e. denser and hotter versus less-dense and cooler). In the next section, we push forward the exploration of the effect of the environment of short and long filaments on the relative distribution of DM, gas, and stars by studying the radial profiles of the baryon fraction around cosmic filaments.

\section{\label{Sect:BaryonFrac}Baryon fractions in filaments}

\subsection{\label{SubSect:Bar_pro}Radial profiles}

    \begin{figure*}
    \centering
   \includegraphics[width=1\textwidth]{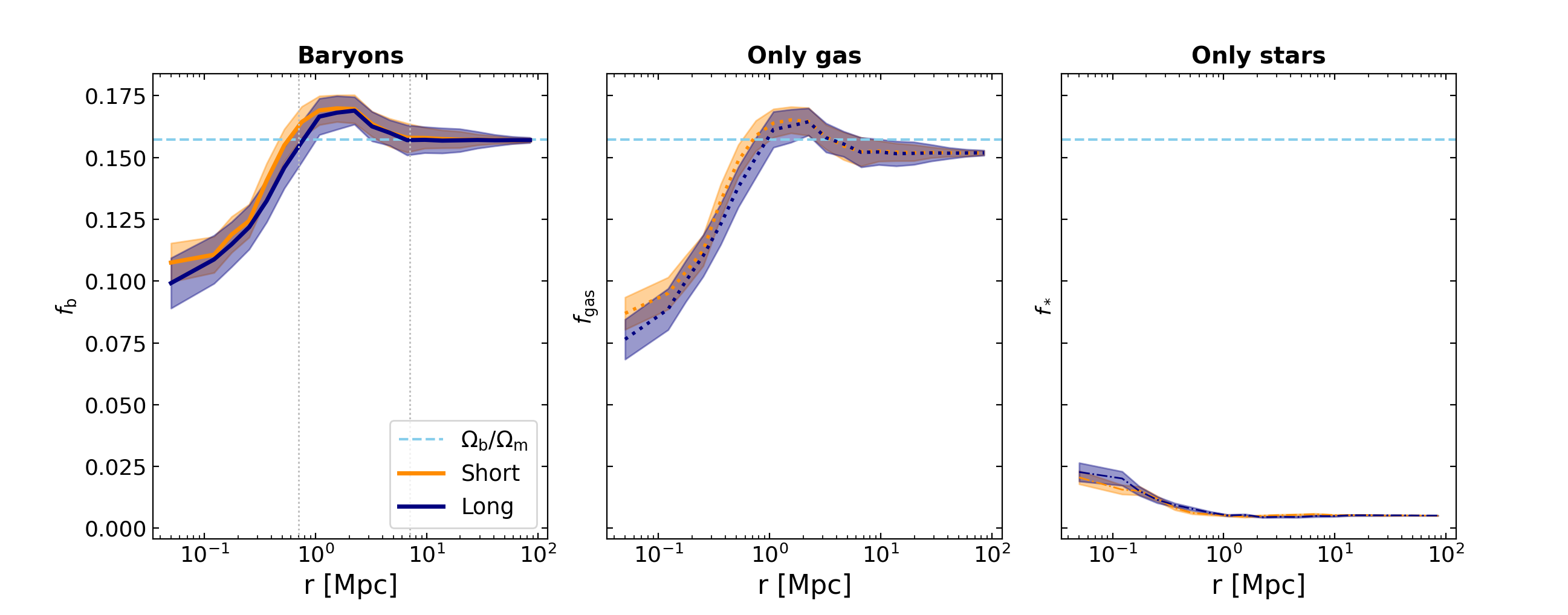}
   \caption{Baryon fraction radial profiles of filaments (first panel), and the individual contributions of gas and stars (second and third panels). The dashed horizontal line shows the cosmic value of $f_\mathrm{cosmic} = \Omega_\mathrm{b} / \Omega_\mathrm{m} = 0.157$.}
    \label{Fig:fraction_b_g_s}
    \end{figure*}

We compute the radial profiles of the baryon fraction in filaments, $f_\mathrm{b}(r)$, by combining the densities presented in Sect.~\ref{SubSect:Densities_absolute} in the following way:
\begin{equation}
    f_\mathrm{b}(r) \equiv \frac{\rho_\mathrm{b}(r)}{\rho_\mathrm{TOT}(r)} = \frac{\rho_\mathrm{gas}(r) + \rho_*(r)}{\rho_\mathrm{DM}(r) + \rho_\mathrm{gas}(r) + \rho_*(r)}.
\end{equation}
The baryon fraction profiles are presented in the first panel of Fig.~\ref{Fig:fraction_b_g_s}. The individual contributions of gas and stars, namely the gas fraction ($f_\mathrm{gas}$) and stellar fraction profiles ($f_*$), are shown in the second and third panels of this figure. At all radii, we have $f_\mathrm{b} = f_\mathrm{gas} + f_*$. \\

As expected, we see that the baryon fraction profiles of filaments converge to the cosmic value, $f_\mathrm{cosmic} = \Omega_\mathrm{b} / \Omega_\mathrm{m} = 0.157$ \citep{Planck2015Cosmo} at very large radii. A deviation from $f_\mathrm{cosmic}$ arises as we approach the cores of filaments. 
The departure from the cosmic value begins at a specific radial scale of $r \sim 7$ Mpc (outer grey vertical line), that is hereafter called $r_\mathrm{ext}$. While at large radial distances of $r > r_\mathrm{ext}$, matter is exclusively ruled by gravity and simply falls into the filament potential wells, at distances $r \leq r_\mathrm{ext}$ baryons stop following the cosmic fraction, showing that their distribution is altered by the mere presence of the filament.

Moreover, it is interesting to note that the $r_\mathrm{ext}$ scale coincides well with the sharp increase of gas temperature in the filament temperature profiles of \cite{GalarragaEspinosa2021}.
All this allows us to interpret $r_\mathrm{ext}$ as the radial extent to which baryonic processes taking place in filaments can affect the large-scale distribution of baryons.
We check this interpretation by computing the total baryon fraction (normalised to the cosmic value) enclosed within a cylinder of radius $r_\mathrm{ext}$ centred on the filament axis. We define this quantity as: 
\begin{equation}\label{Eq:F_rext}\displaystyle
    F(r<r_\mathrm{ext}) \equiv \frac{\displaystyle f_\mathrm{b}(r<r_\mathrm{ext})}{f_\mathrm{cosmic}} = \frac{\displaystyle \int_0^{r_\mathrm{ext}} r\, \left[\rho_\mathrm{gas}(r) + \rho_*(r) \right] \, dr}{f_\mathrm{cosmic} \times \displaystyle \int_0^{r_\mathrm{ext}} r\, \rho_\mathrm{TOT}(r) \, dr},
\end{equation}
and the resulting values for the average short and long filaments are respectively $1.02^{+0.05}_{-0.05}$ and $1.01^{+0.06}_{-0.05}$, i.e. compatible with one.

Remarkably, short and long filaments exhibit the same value of $r_\mathrm{ext} \sim 7$ Mpc and baryon fraction profiles that significantly overlap. In line with the results of Sect.~\ref{SubSect:Fit_OV}, this means that the distribution of gas and stars with respect to that of DM is rather independent of the specific absolute density ranges (Fig.~\ref{Fig:density_profiles}) and large-scale environments particular to each filament population. Potentially, this similarity between filament populations suggests that the baryonic processes responsible for modifying the baryonic density profile relative to that of the DM are mostly independent of the environment.\\


At radii $r \leq r_\mathrm{ext}$, i.e. inside the filaments, the baryon fraction profiles show two very different regimes, whose transition happens at a radius of $\sim 0.7$ Mpc. This radius is hereafter called $r_\mathrm{int}$ and is identified as the intersection point between $f_\mathrm{b}$ and $\Omega_\mathrm{b} / \Omega_\mathrm{m}$ (inner grey line in Fig.~\ref{Fig:fraction_b_g_s}).
We note that this distance is also consistent with the radius where the DM and gas over-density profiles of Fig.~\ref{Fig:density_FILS_RESCALED} separate. 
The two regimes in the baryon fraction profiles are identified in the following, from the inner to the outer regions. 
Firstly, a clear baryon depletion is seen at the cores, as shown by the sharp decrease of the baryon fractions. This depletion is quantified by the factor $Y_\mathrm{b}(r<r_\mathrm{int}) \equiv f_\mathrm{b}(r<r_\mathrm{int})/f_\mathrm{cosmic}$, which respectively gives $0.90^{+0.06}_{-0.06}$ and $0.85^{+0.08}_{-0.07}$ in short and long filaments, consistently with the results of \citet[][we note the similar values between the two different filament populations]{GhellerVazza2019_surveyTandNTprops_fils}. For comparison, the depletion factor in filaments is of the same order of magnitude than the usual value in cores of clusters of galaxies, $Y_\mathrm{b} \sim 0.85$ \citep[e.g.][]{Kravtsov2005, Planelles2013}.
Secondly, an excess of baryons with respect to the cosmic value, characterised by a bump in the profiles, is observed.
These two different regimes can be understood by focusing only on the gas component. Indeed the fraction of gas $f_\mathrm{gas}$ in filaments (second panel of Fig.~\ref{Fig:fraction_b_g_s}) shows the same features as the total fraction of baryons, i.e. a depletion at the cores ($r < r_\mathrm{int}$), and a bump in the $[r_\mathrm{int}, r_\mathrm{ext}]$ radial range. This resemblance between the total fraction and that of gas is not surprising, given that gas represents on average $\sim 78 \%$ of the baryonic mass in the innermost bin, and $\sim 96\%$ in  the $[r_\mathrm{int}, r_\mathrm{ext}]$ interval, thus corresponding to the most abundant form of baryonic matter in filaments.
Sections~\ref{SubSect:baryon_depletion} and \ref{SubSect:Bump} will respectively present the analysis of the gas depletion and the bump with respect to the background value.\\

Before analysing the two regimes identified above, let us comment on the stellar fractions of filaments, $f_*$ (the third panel of Fig.~\ref{Fig:fraction_b_g_s}). These show a flat trend in the profiles at large distances from the spines of filaments, and contrary to gas, $f_*$ increases at the filament cores ($r < r_\mathrm{int}$). This increase of the fraction of stars at the cores reflects well the results of previous studies of galaxies which have found that cores of filaments are more populated by massive galaxies than more remote regions \citep{Malavasi2017, Laigle2016cosmos, Kraljic2018, Kraljic2019, Bonjean2020filaments}.

\subsection{\label{SubSect:baryon_depletion}Analysis of the gas depletion at $r < r_\mathrm{int}$}

Gas depletion can be easily understood by analogy with the case of clusters of galaxies \citep[e.g.][and references therein]{Kravtsov2005, Puchwein2008, Planelles2013, LeBrun2014, Barnes2017_EAGLEclusters, McCarthy2017_BAHAMAS, Chiu2018_SPTclusters, Henden2020}. In these structures, the decrease of the gas fraction is due to the conversion of gas into stars (that lowers the gas fraction and enhances the stellar one), and the ejection of gas due to feedback events \cite[which displace, disperse, and redistribute the gas, e.g.][]{CenOstriker2006}. 
Previous studies have shown that  feedback effects from active galactic nuclei (AGNs) can modify the distribution of baryons up to several Mpc away from these objects \citep[see e.g.][]{Chisari2018}. Thus, in this section we investigate whether the feedback from AGNs sitting in filaments is potentially powerful enough to deplete the filament cores by ejecting their gas away.
A fundamental requirement for filamentary gas ejection is that the injected energy (coming from the feedback of AGNs inside filaments) be higher than the gas gravitational potential energy (i.e. competition between pushing and pulling of gas).\\

Firstly, we compute the gravitational potential $\phi$ felt by a test particle at a distance $R$ from the spine of the filament using the following expression (see Appendix~\ref{Appendix:POTENTIALS} for all the details):
\begin{equation}\label{Eq:Phi_fils}
    \phi(R) = 2 \, \mathrm{G} \int_0^R \mathrm{d}r \, \frac{1}{r} \int_0^r \mathrm{d}r' r' \Delta \rho_\mathrm{TOT}(r') + \phi_0.
\end{equation}
In this equation, $G$ is the universal gravitational constant, $\Delta \rho_\mathrm{TOT}$ is the total matter over-density ($\Delta \rho_\mathrm{TOT} = \rho_\mathrm{TOT} - \rho_\mathrm{bkg}$), and $\phi_0$ is an integration constant chosen so that the potential at the limit between the interior and the exterior of the filament is zero, i.e. $\phi(R_\mathrm{lim}) = 0$. Thus, Eq.~\ref{Eq:Phi_fils} gives the following value of $\phi_0$:
\begin{equation}
    \phi_0 = - 2 \, \mathrm{G} \int_0^{R_\mathrm{lim}} \mathrm{d}r \, \frac{1}{r} \int_0^r \mathrm{d}r' r' \Delta \rho_\mathrm{TOT}(r').
\end{equation}
We note that, based on the studies of the baryon fractions profiles presented above (Fig.~\ref{Fig:fraction_b_g_s}), the limit between the interior and the exterior of the filament is set to $R_\mathrm{lim} = r_\mathrm{ext} = 7$ Mpc. 

\begin{figure}
    \centering
   \includegraphics[width=0.45\textwidth]{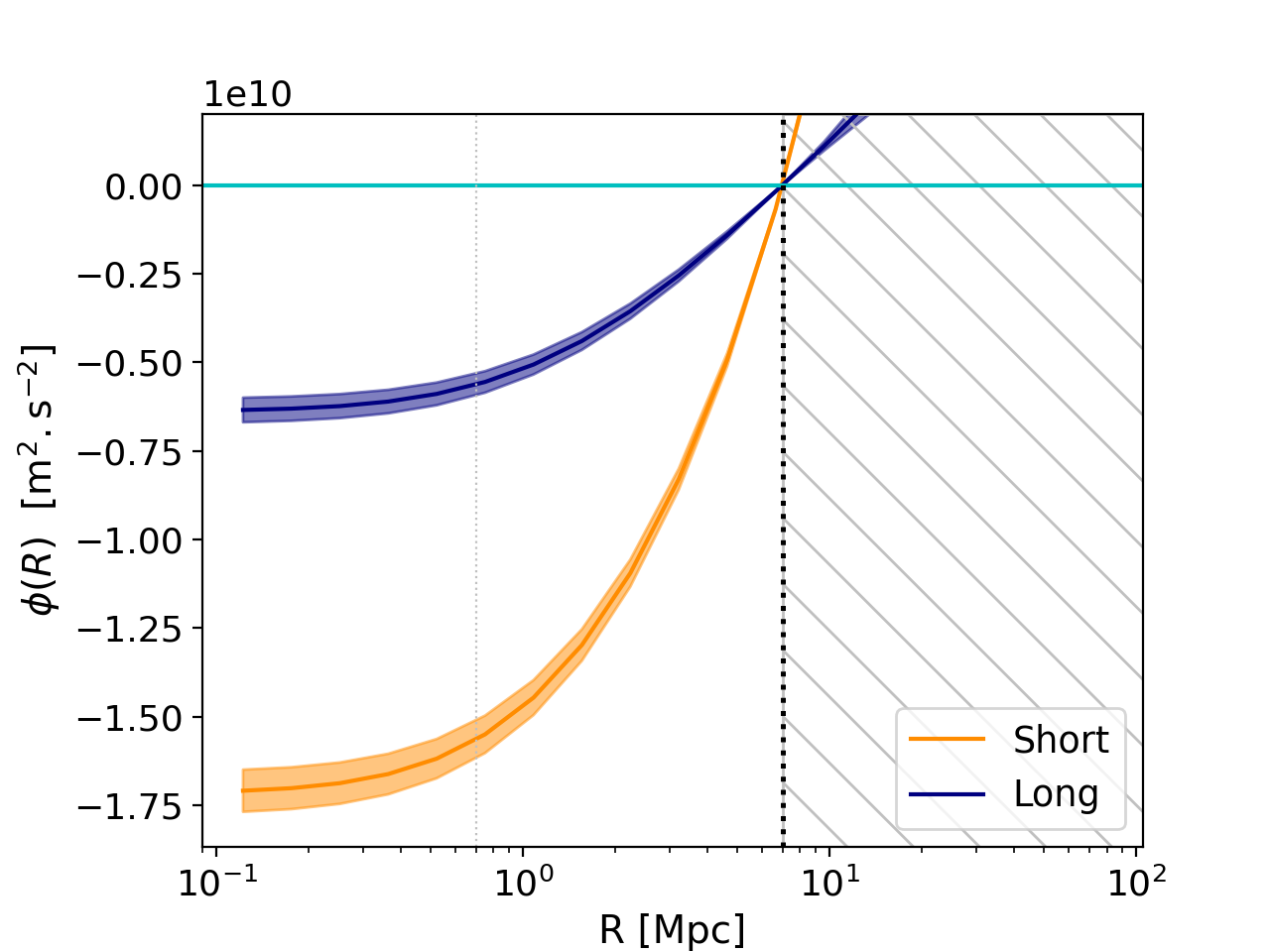}
   \caption{Gravitational potential (Eq.~\ref{Eq:Phi_fils}) felt by a test mass inside short (orange) and long filaments (blue). The limit between inside and outside for the computation of the gravitational potential is set to $R_\mathrm{lim} = r_\mathrm{ext} = 7$ Mpc. The outside regions are hatched in grey.}
    \label{Fig:gravitational_potential}
    \end{figure}

The resulting $\phi(R)$ profiles for short and long filaments are shown in Fig.~\ref{Fig:gravitational_potential}. As expected, the gravitational potential of short filaments is deeper than that of the long filaments. This is due to the higher density values of the former with respect to the later, e.g.~three times higher at the cores (see Fig.~\ref{Fig:density_profiles}). We note that this factor of three between the densities of short and long filament cores is also reflected in the values of the gravitational potential.
Physically, the calculations presented above show that, in order to be displaced outside of filaments, baryonic gas particles need to acquire a kinetic energy that is bigger than their potential energy in the depth of the gravitational well
\begin{equation}\label{Eq:DeltaPhi}
    |\Delta \phi| \equiv |\phi(0) - \phi(R_\mathrm{lim})| = | \phi_0 |.
\end{equation}\\
By construction, the values of $|\Delta \phi|$ can by read directly from the first bin of Fig.~\ref{Fig:gravitational_potential}.\\

    \begin{figure}
    \centering
   \includegraphics[width=0.45\textwidth]{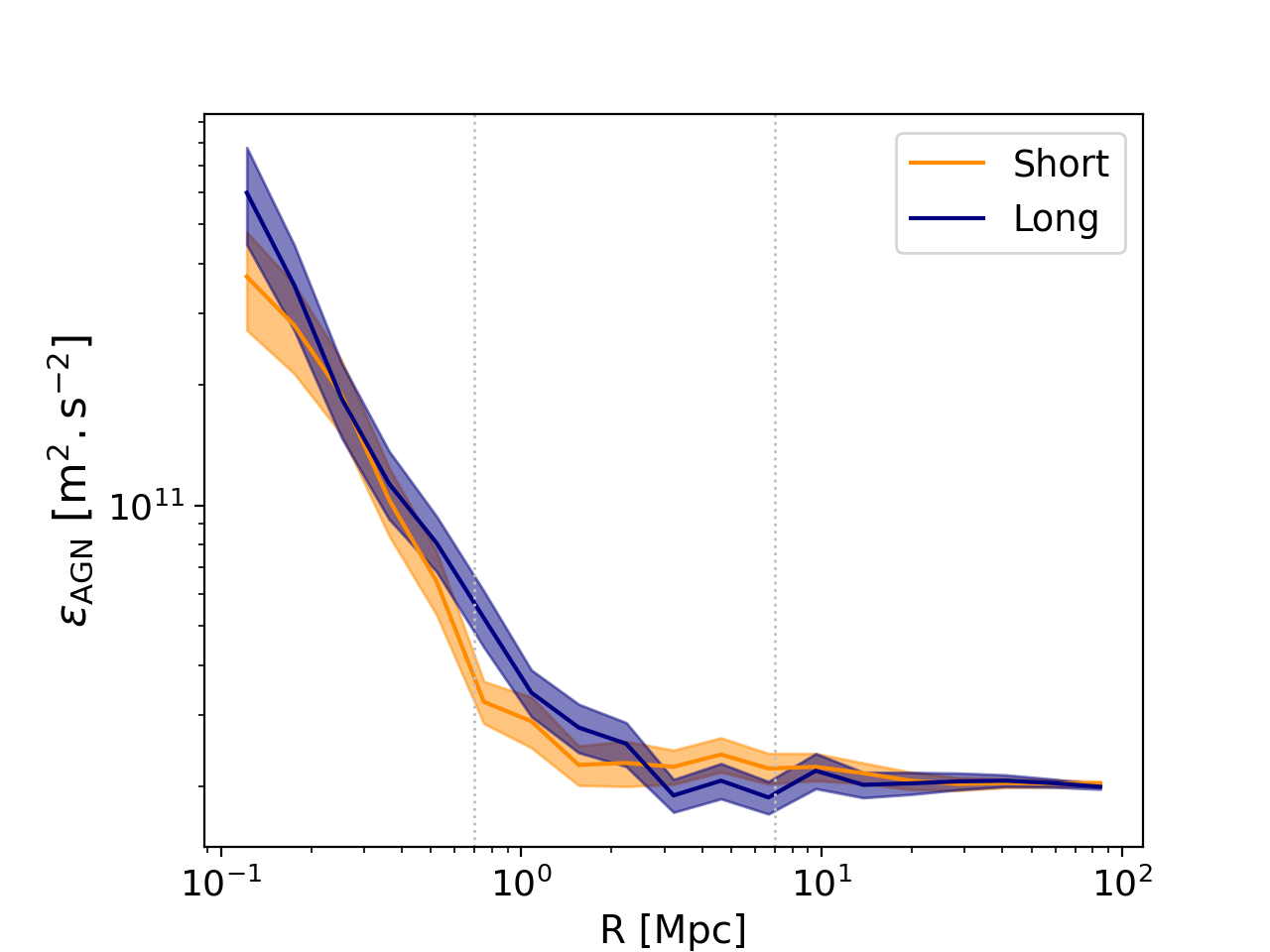}
   \caption{Radial profiles of the kinetic energy by unit of gas mass injected by AGNs into the medium (see Eq.~\ref{Eq:Eagn}). The inner and outer vertical dashed lines correspond respectively to the $r_\mathrm{int}$ and $r_\mathrm{ext}$ radii.}
    \label{Fig:E_AGN_profiles}
    \end{figure}
    
    \begin{figure}
    \centering
   \includegraphics[width=0.45\textwidth]{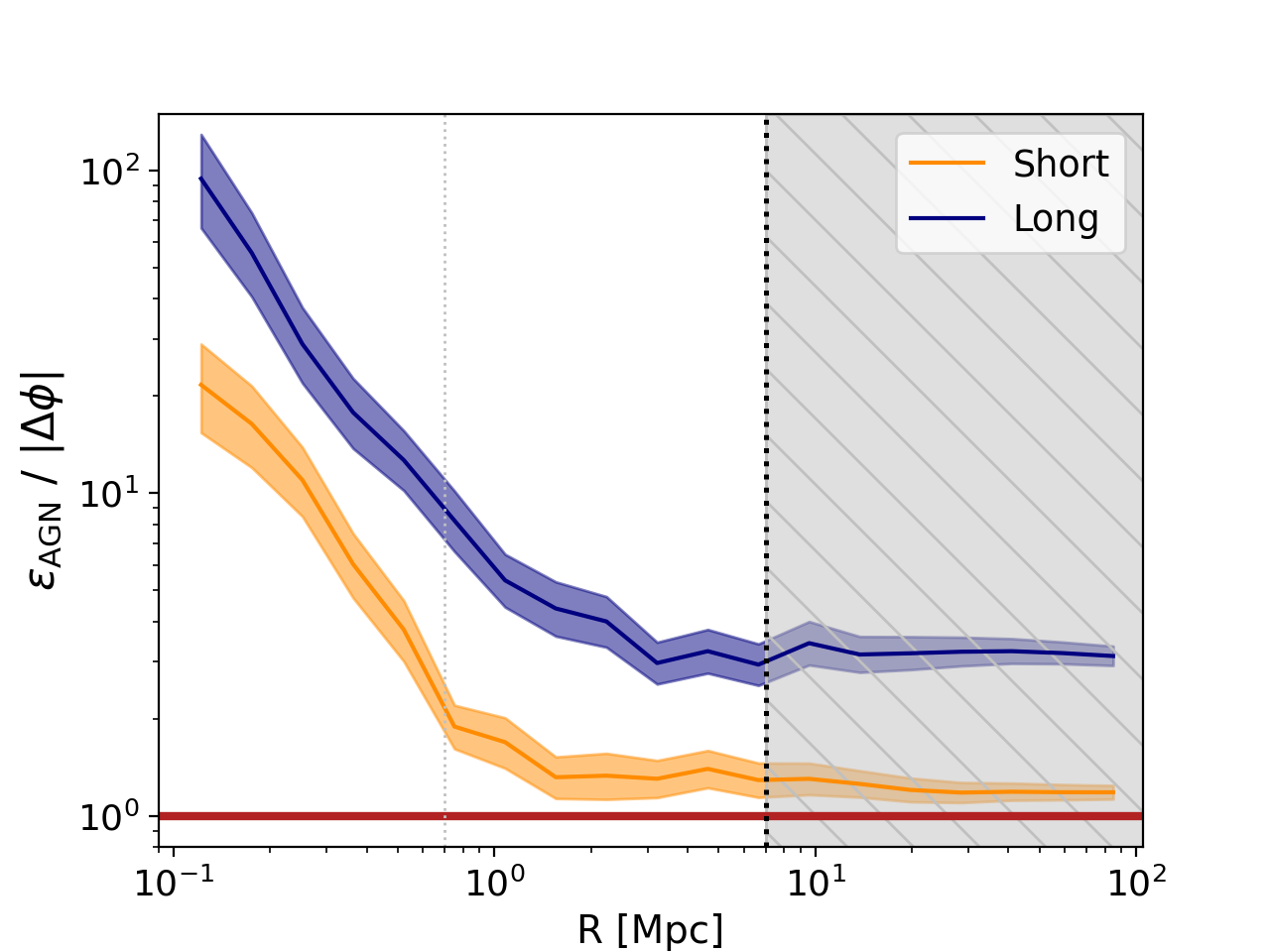}
   \caption{Ratio between the AGN feedback energy by unit of gas mass, $\varepsilon_\mathrm{AGN}$ (Eq.~\ref{Eq:Eagn}), and the depth of the gravitational potential, $|\Delta \phi|$ (Eq.~\ref{Eq:DeltaPhi}). The inner and outer vertical dashed lines correspond respectively to the $r_\mathrm{int}$ and $r_\mathrm{ext}$ radii. The regions outside filaments (i.e. $r > r_\mathrm{ext}$) are hatched in grey and must not be considered in the analysis.}
    \label{Fig:E_RATIO}
    \end{figure}

Let us now focus on the energy injected by AGNs. We compute the amount of energy by unit of gas mass, $\varepsilon_\mathrm{AGN}$, injected by AGN feedback events into a cylindrical shell of thickness $|r_{k-1} - r_{k}|$ around the axis of the filament using the following equation (see Appendix~\ref{Appendix:POTENTIALS} for details):
\begin{equation}\label{Eq:Eagn}
    \varepsilon_\mathrm{AGN}(R) = \frac{\int_{r_{k-1}}^{r_{k}} \mathrm{d}r \,r \,\, e_\mathrm{AGN}^\mathrm{kin}(r) }{ \int_{r_{k-1}}^{r_{k}} \mathrm{d}r \, r \,\, \rho_\mathrm{gas}(r)},
\end{equation}
where $R \in [r_{k-1}, r_{k}]$. Here $e_\mathrm{AGN}^\mathrm{kin}$ is the energy per unit volume, which corresponds to the cumulative amount of kinetic AGN feedback energy (total over the entire lifetime of the corresponding black-holes) injected into the surrounding gas, divided by the volume of the cylinder centred on the filament axis. For reference, the $e_\mathrm{AGN}^\mathrm{kin}$ profiles are presented in Fig.~\ref{Fig:APP_eAGN_profiles}.
We note that, from the two possible AGN energy injection modes available in the TNG model (i.e. thermal and kinetic), only  the kinetic mode (field \texttt{BHCumEgyInjectionRM} in the PartType5 catalogue) is studied here. It was indeed shown that it is the only mode able to expel gas away from the centre of galaxies \citep{Weinberger2018_AGN, Terrazas2020_AGN, Quai2021_AGN} thanks to the release of mechanical energy into the gas (via winds in random directions). On the other hand, as shown by \cite{Weinberger2018_AGN}, the AGN thermal injection mode does not efficiently couple with the gas in the galaxy environment, and thus it does not alter its spatial distribution \citep[see also][]{Terrazas2020_AGN, Quai2021_AGN}. Of course, this thermal injection mode contributes to heating the gas in the inter-galactic medium in the filament.

The $\varepsilon_\mathrm{AGN}$ profiles resulting from Eq.~\ref{Eq:Eagn} are presented in Fig.~\ref{Fig:E_AGN_profiles}.
Overall, the two filament populations have very similar profiles. This is due to the fact that the differences between short and long filaments shown by the $e_\mathrm{AGN}^\mathrm{kin}$ profiles (Fig.~\ref{Fig:APP_eAGN_profiles}) are compensated by the different gas masses of the two populations. This means that the potential efficiency of AGN feedback events relative to the gas distribution is quite the same in short and long filaments. We note, however, the slightly higher $\varepsilon_\mathrm{AGN}$ value at the inner cores of long filaments.\\

Let us finally investigate whether these feedback processes are potentially powerful enough to contribute to the gas depletion observed at the cores of filaments (see Fig.~\ref{Fig:fraction_b_g_s}).
This is done by comparing the values of the injected AGN energy ($\varepsilon_\mathrm{AGN}$, Eq.~\ref{Eq:Eagn}) to the depth of the gravitational potential well ($|\Delta \phi|$, shown in Eq.~\ref{Eq:DeltaPhi}).
The ratio between these two quantities, $\varepsilon_\mathrm{AGN} / |\Delta \phi|$ is presented in Fig.~\ref{Fig:E_RATIO}. 
For both short and long populations, we find that the ratio $\varepsilon_\mathrm{AGN} / |\Delta \phi|$ is larger than one at all radii, reaching the maximum values of respectively $\sim 21$ and $\sim 94$ at the filament cores. This means that, within $r< r_\mathrm{ext}$ Mpc, the gas thrust due to AGN feedback is theoretically strong enough to counteract the gravitational pull towards the filament cores. Of course, it is important to remember that not all the kinetic energy deposited in the medium by AGN feedback processes is necessarily transferred to the gas. Indeed, this depends on several factors such as the AGN jet geometry and orientation, their stability, and the coupling between the energetic jets and winds and the gas, which is not the focus of this study. The main result of this study is that the feedback from AGNs can potentially inject enough energy into the medium to contribute to the gas depletion observed in Fig.~\ref{Fig:fraction_b_g_s}. 

Figure~\ref{Fig:E_RATIO} also shows that the efficiency of this process depends on the filament population. Indeed, due to their shallower gravitational potential, long filaments can be more easily affected by AGN feedback processes than the short population, as shown by the higher values of $\varepsilon_\mathrm{AGN} / |\Delta \phi|$ at all radii of the former with respect to the latter. This can be interpreted in analogy with clusters of galaxies, where the link between gas depletion and density has been thoroughly studied, finding that AGN feedback effects are efficient enough to displace large amounts of gas outside the potential wells of the small-scale structures (groups), but barely affect the more massive (larger-scale) clusters of galaxies \citep[e.g.][and references therein]{Planelles2013, LeBrun2014, McCarthy2017_BAHAMAS, Barnes2017_EAGLEclusters, Henden2020}.\\

Naturally, one might think that the observed baryon depletion discussed in this section could be solely the result of the thermal pressure of filament gas. Nevertheless, by comparing the outputs between simulations with and without the inclusion of AGN feedback, \cite{GhellerVazza2019_surveyTandNTprops_fils} have shown that AGNs effectively affect the baryon fraction distribution of filaments. In that paper, the baryon fraction $f_b$ enclosed in filaments in the simulations that include AGN feedback reaches values that are smaller than those of the structures detected in the baseline simulation without feedback. In line with the results presented in this section, this supports a picture in which the effects of AGNs have a non-negligible impact on the distribution of baryons in large-scale filaments.

\subsection{\label{SubSect:Bump}Analysis of the bump at $r \in [r_\mathrm{int}, r_\mathrm{ext}]$}

    \begin{figure*}
    \centering
   \includegraphics[width=1\textwidth]{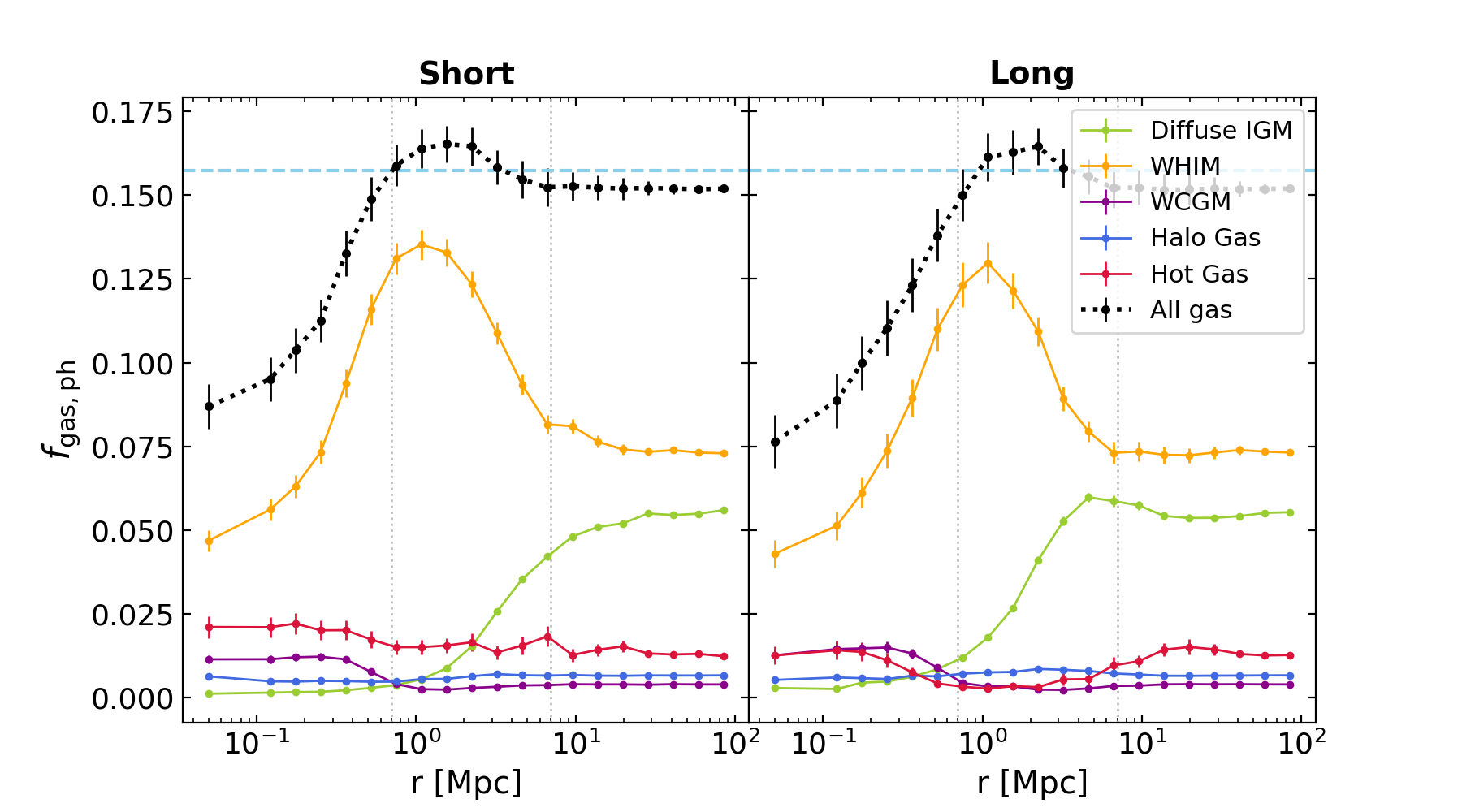}
   \caption{Gas fraction profiles separated in five different phases, according to the classification of \citep{Martizzi2019a}. The vertical lines show the positions of the $r_\mathrm{int}$ (left) and $r_\mathrm{ext}$ (right) radial scales.}
    \label{Fig:FG_gas_phases}
    \end{figure*}
    
We now focus on the bump of the profiles, observed at $r \in [r_\mathrm{int}, r_\mathrm{ext}]$ in Fig.~\ref{Fig:fraction_b_g_s}.
In order to understand this feature, it is primordial to disentangle between the different `types' of gas that co-exist in the cosmic web, e.g. the diffuse inter-galactic medium, the warm-hot intergalactic medium (WHIM), hot gas, etc. Based on their position in the density versus temperature phase-space, we separate gas into the different phases as defined by \cite{Martizzi2019a} and we run the same analysis for each of them. The resulting gas fractions are shown in Fig.~\ref{Fig:FG_gas_phases}, where it is evident that the observed bump is specific to gas in the WHIM, with temperature $T \sim 10^5-10^7$ K (warm), and density $n_\mathrm{H} < 10^{-4} \, \mathrm{cm}^{-3}$ (diffuse). 
Indeed, diffuse gas is accreted towards the cores of filaments by gravitational attraction, and due to its collisional nature, it is shock heated and converted into WHIM gas. Unable to efficiently cool down, WHIM gas assembles at the outskirts ($r \sim 1$ Mpc) of filaments. We note that these results are in excellent agreement with the study of gas properties around filaments \citep{GalarragaEspinosa2021}.

Interestingly, the work of \cite{Rost2021}, focusing on the filaments at the outskirts of clusters and their connection to these structures, has also detected a bump in the gas fraction profiles of filaments connected to the clusters of \textsc{The ThreeHundred} simulation \citep{Cui2018_ThreeHundred}. In that work, this feature was exclusively interpreted as the turbulent fuelling of clusters with filament gas (i.e. filaments acting as highways of gas towards clusters). However, since the outskirts of clusters are not included in the present analysis (regions within $3 \times R_{200}$ from cluster centres are excluded), the bump highlighted here is rather found to be an intrinsic property of radial accretion shocks of WHIM gas at the boundaries of filaments.
Of course, this intrinsic radial feature can be enhanced by other processes,
like the turbulent motions for the filamentary structures at the outskirts of clusters detected by \cite{Rost2021}.


\section{\label{Sect:Scalings}Scaling relations with gas temperature and pressure}

    \begin{figure}
    \centering
   \includegraphics[width=0.45\textwidth]{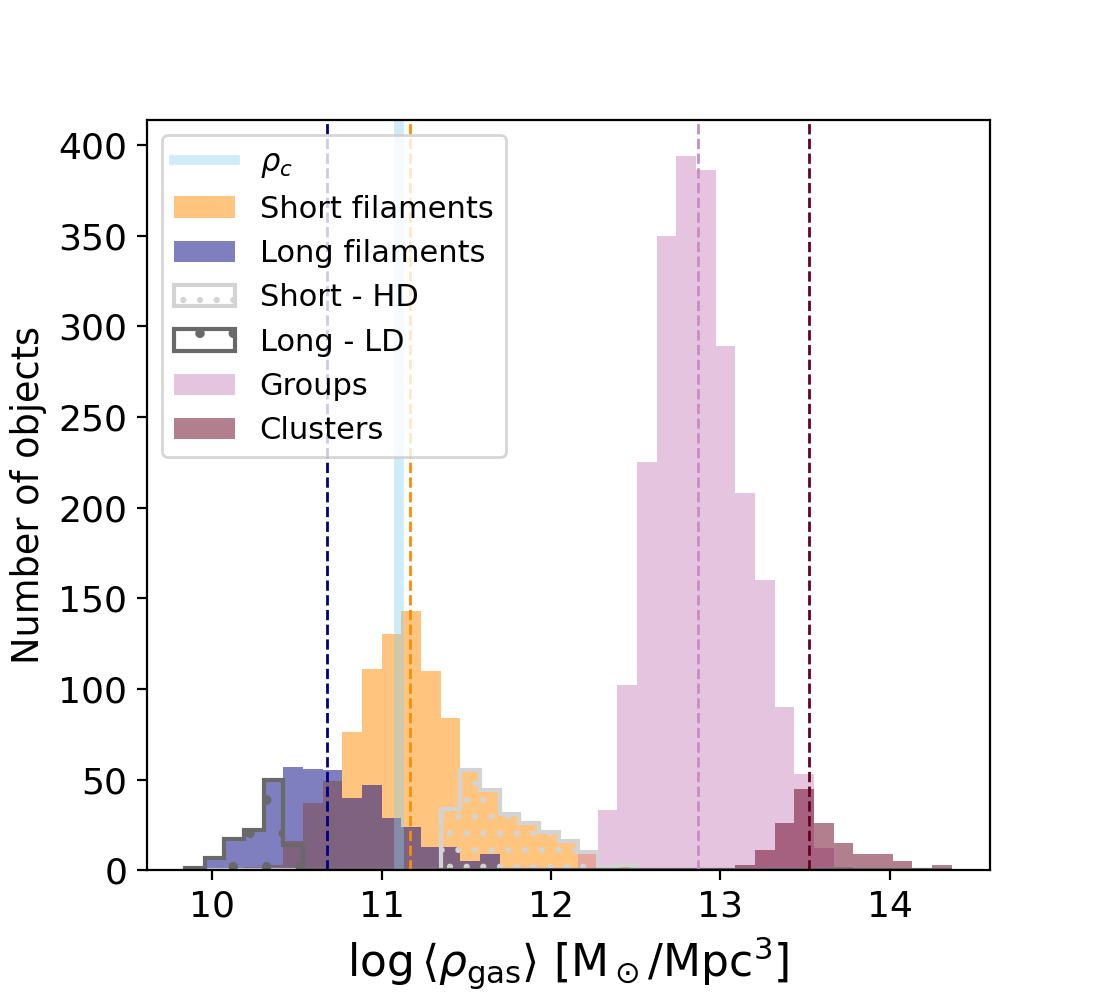}
   \includegraphics[width=0.45\textwidth]{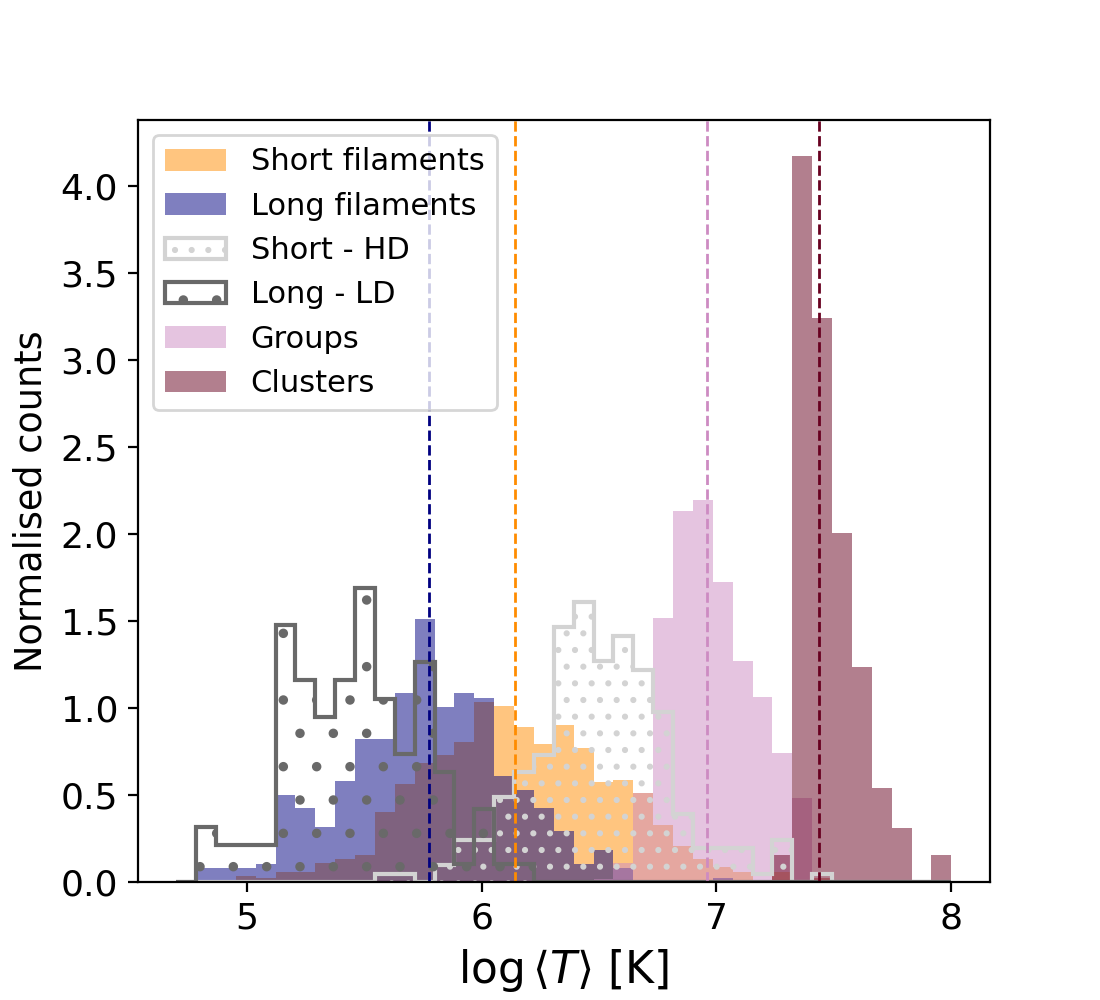}
   \includegraphics[width=0.45\textwidth]{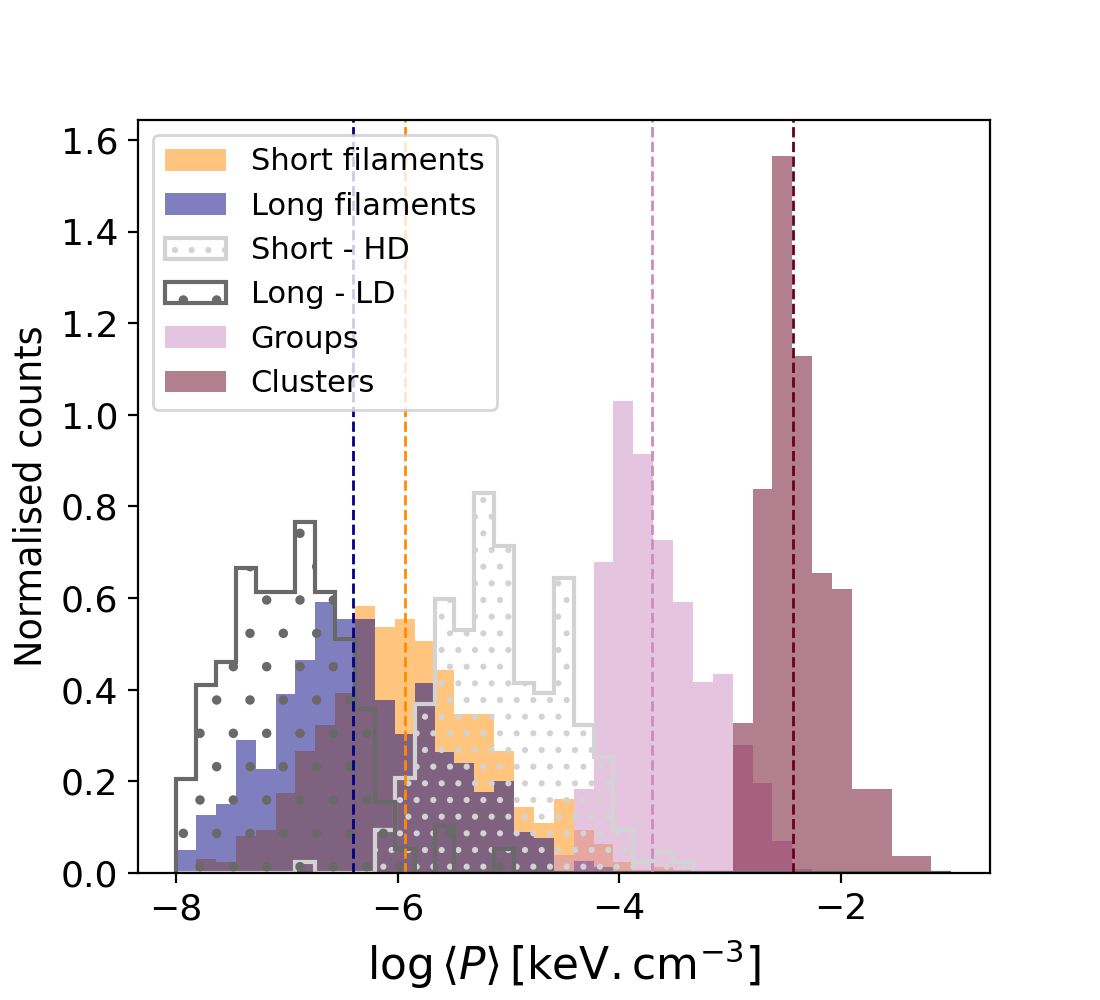}
   \caption{$\langle \rho_\mathrm{gas} \rangle$,  $\langle T \rangle$, and $\langle P \rangle$ distributions of filaments and clusters. The vertical dashed lines correspond to the median values of each distribution. For reference, the critical density of the Universe, $\rho_c$, is presented in the first panel by the vertical light-blue line.}
    \label{Fig:rho_T_P_distributions}
    \end{figure}
    
After analysing the full baryonic content of cosmic filaments, let us now focus our analysis only on the gaseous component, whose observation is the most challenging nowadays. Indeed, despite recent detection of the SZ effect around filaments (see e.g. \citet{DeGraaff2019} and \citet{Tanimura2020_byopic} results), and studies of X-ray emission of filament gas (see e.g. \citet{Tanimura2020_Rosat} and \citet{Biffi2021_eROSITAfils} respectively in ROSAT and in the very recent eROSITA data), the gaseous content of cosmic filaments is still not fully constrained in observations. In order to help future studies, we provide scaling relations between gas density, temperature, and pressure, derived from the analysis of the TNG300-1 simulation.

From the gas density profile of each filament, we compute the mean gas density (hereafter $\langle \rho_\mathrm{gas} \rangle$) up to the radial scale $r_\mathrm{ext}$, defined as the radial extension of baryons in filaments (see Fig.~\ref{Fig:fraction_b_g_s}). The same is done for gas temperature and pressure, whose radial profiles were computed following Eq.~6 and 7 of \cite{GalarragaEspinosa2021}. 
This yields one $\langle \rho_\mathrm{gas} \rangle$, $\langle T \rangle$ and $\langle P \rangle$ value for each filament of our catalogue. 
We note that, in order to reduce the scatter, in this section we focus only on the filaments whose profiles are complete (i.e. without empty bins). The resulting number of filaments in the short and long populations is respectively $969$ and $446$ (for reference, the initial number in each population is $2846$ and $632$).

With the aim of comparing the scaling relations of filaments to those of other cosmic structures, in this section we also show the $\rho_\mathrm{gas}-T$ and $\rho_\mathrm{gas}-P$ relations of clusters of galaxies, the densest structures of the cosmic web. These structures were selected from the publicly available friend-of-friend (FoF) halo catalogue of the TNG300-1 simulation box. We focus on the most massive haloes of this catalogue, that we separate into three classes according to the mass: $13 \leq \log M_{200} < 14$ (2369 haloes), $14 \leq \log M_{200} < 14.5$ (138), and $\log M_{200} \geq 14.5$ $\mathrm{M}_\odot /h$ (15 haloes), corresponding respectively to groups of galaxies, clusters, and massive clusters. For further details on the FoF group catalogue, we refer the reader to \cite{Nelson2019_TNGdata_release}. In the case of clusters, the corresponding average quantities are computed up to the $R_{200}$ radial scale.

The $\langle \rho_\mathrm{gas} \rangle$,  $\langle T \rangle$, and $\langle P \rangle$ distributions of filaments and clusters are presented in Fig.~\ref{Fig:rho_T_P_distributions}.
For comparison, this figure also shows the specific distributions of extreme filaments, i.e. the high-density (HD) short, and low-density (LD) long filaments. These filaments correspond respectively to the 25 and 75 percentiles of the short and long populations, and account for a number of 242 (short HD) and 112 (long LD). The distributions of Fig.~\ref{Fig:rho_T_P_distributions} highlight the hierarchical structure of the cosmic web, with the densest structures (i.e. clusters) being hotter and with higher pressure values than the least dense ones (i.e.\ long filaments).\\

    \begin{figure*}
    \centering
   \includegraphics[width=0.5\textwidth]{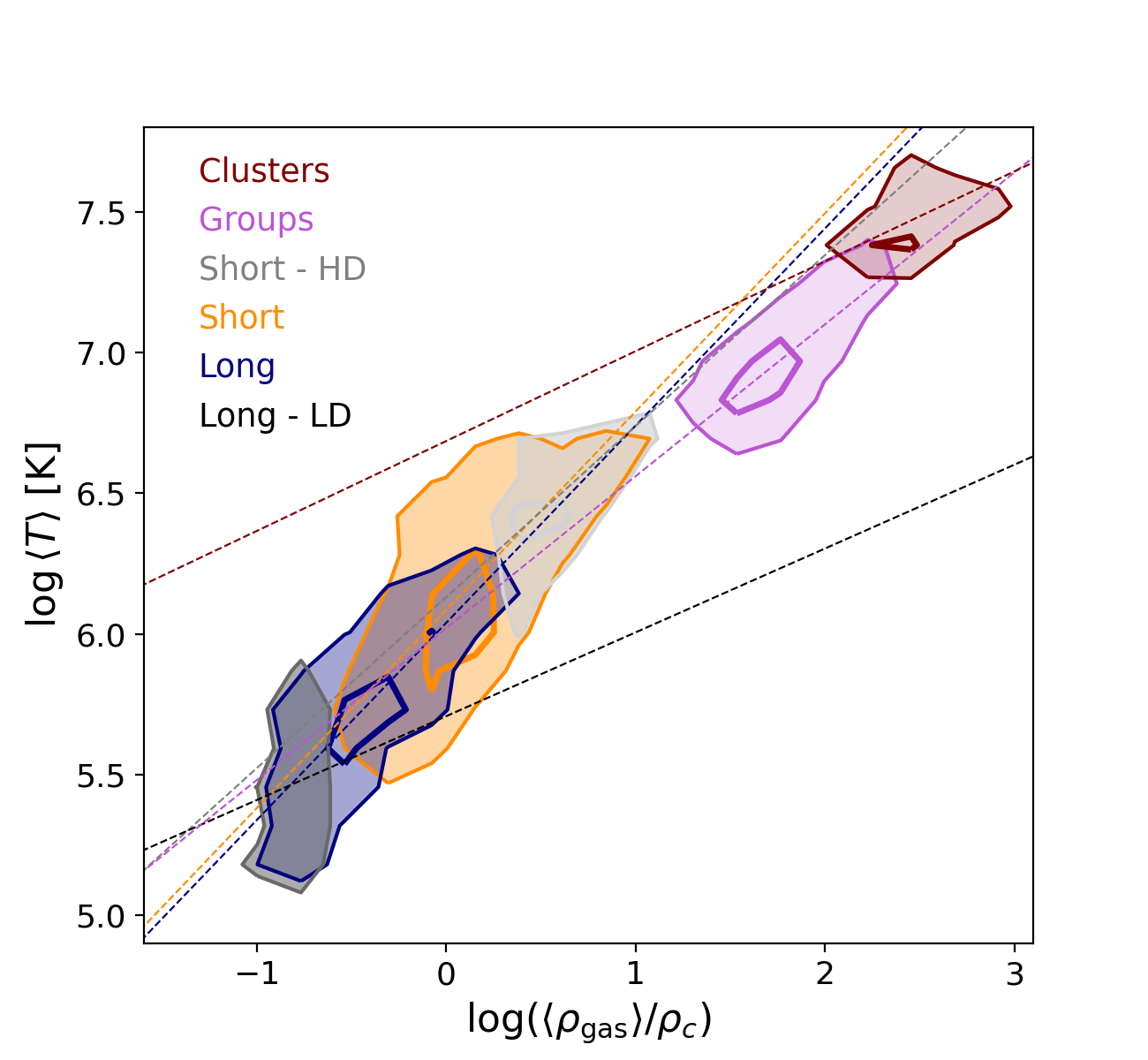}\includegraphics[width=0.5\textwidth]{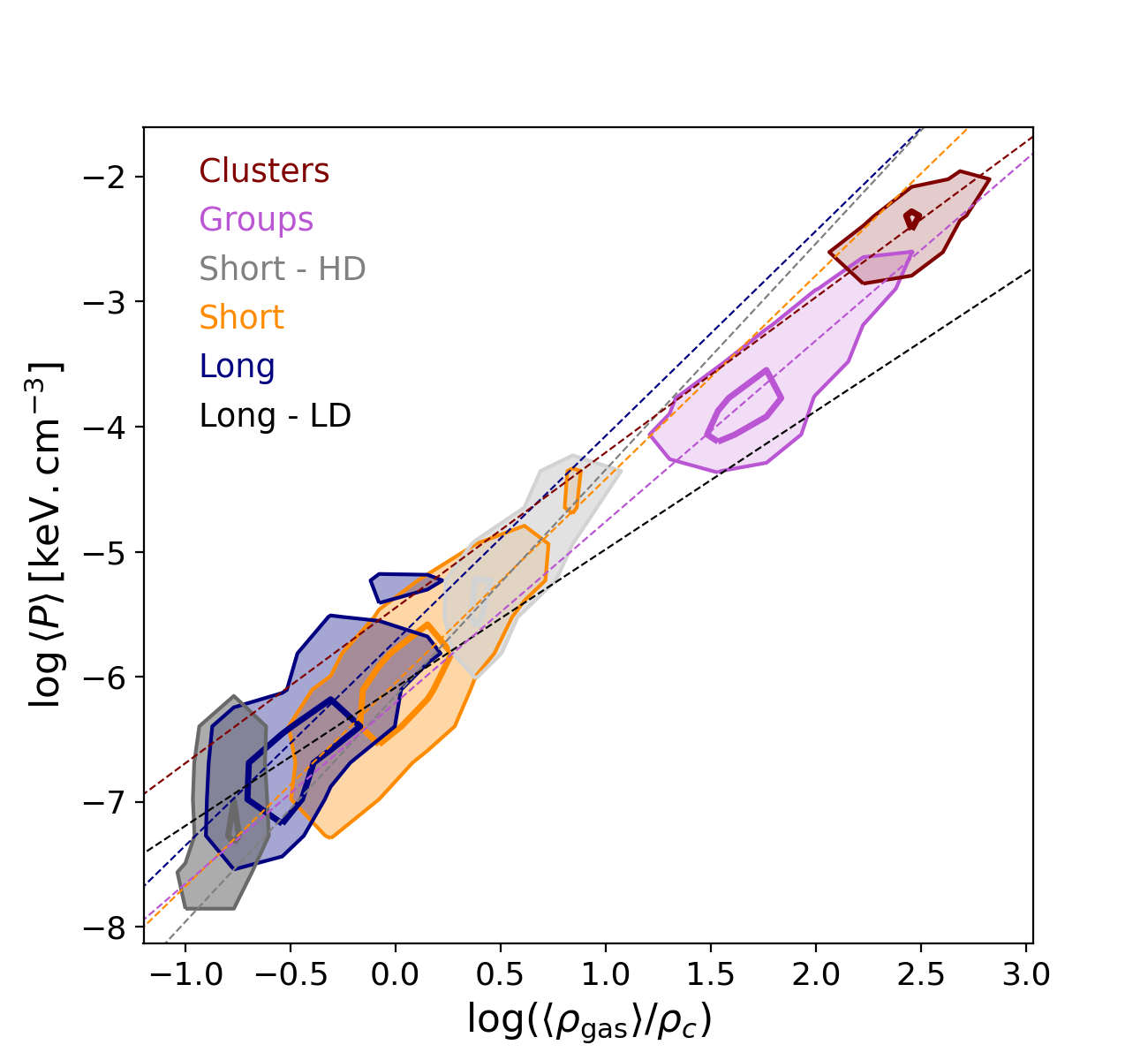}
   \caption{Scaling relations between temperature (left) and pressure (right) with gas density, for filaments and galaxy clusters. For each structure, the dashed line corresponds to the linear function of Eq.~\ref{Eq:fit_eq_T}, with the parameters of Table~\ref{Table:FIT_T} and \ref{Table:FIT_P} for temperature and pressure, respectively.}
    \label{Fig:Scalings}
    \end{figure*}
    
The scaling relations between gas density and temperature for filaments and clusters are obtained by fitting the following function in the $\langle \rho_\mathrm{gas} \rangle - \langle T \rangle$ plane:
\begin{equation}\label{Eq:fit_eq_T}
   \log \, \langle T \rangle = \mathrm{a} \, \log \left( \langle \rho_\mathrm{gas} \rangle / \rho_c \right) + \mathrm{b},
\end{equation}
where $\rho_c$ corresponds to the critical density of the Universe (see vertical light-blue line of Fig.~\ref{Fig:rho_T_P_distributions}).
The same function form is used for pressure, in the $\langle \rho_\mathrm{gas} \rangle - \langle P \rangle$ plane.

    \begin{table*}
\caption{Results of the scaling relations between pressure and gas density, $\log \, \langle T \rangle = \mathrm{a} \, \log \left( \langle \rho_\mathrm{gas} \rangle / \rho_c \right) + \mathrm{b}$.}
\label{Table:FIT_T}     
\centering  
\begin{tabular}{ c | c c c }
 \hline\hline  
    & $\mathrm{a}$ & $\mathrm{b}$ & $\chi^2_\nu$ \\ \hline
    Clusters & $0.32 \pm 0.03$ & $6.69 \pm 0.08$ & $0.01$ \\
    Groups & $0.54\pm 0.01$ & $6.02 \pm 0.02$ & $0.01$\\
    Short filaments - HD & $0.61 \pm 0.07$ & $6.13 \pm 0.05$ & $0.07$ \\
    Short filaments (all) & $0.70 \pm 0.03$ & $6.09 \pm 0.01$ & $0.10$ \\
    Long filaments (all) & $0.70 \pm 0.03$ & $6.04 \pm 0.02$ & $0.06$ \\
    Long filaments - LD & $0.30 \pm 0.20$ & $5.71 \pm 0.17$ & $0.08$ \\
\hline
 \end{tabular}
\end{table*}

\begin{table*}
\caption{Results of the scaling relations between pressure and gas density, $\log \, \langle P \rangle = \mathrm{a} \, \log \left( \langle \rho_\mathrm{gas} \rangle / \rho_c \right) + \mathrm{b}$.}
\label{Table:FIT_P}     
\centering  
\begin{tabular}{ c | c c c }
 \hline\hline  
    & $\mathrm{a}$ & $\mathrm{b}$ & $\chi^2_\nu$ \\ \hline
    Clusters & $1.24 \pm 0.03$ & $-5.45 \pm 0.08$ & $0.01$\\
    Groups & $1.45 \pm 0.02$ & $-6.21 \pm 0.03$ & $0.05$\\
    Short filaments - HD & $1.81 \pm 0.09$ & $-6.15 \pm 0.06$ & $0.11$ \\
    Short filaments (all) & $1.63 \pm 0.04$ & $-6.05 \pm 0.01$ & $0.19$ \\
    Long filaments (all) & $1.69 \pm 0.06$ & $-5.71 \pm 0.03$ & $0.22$\\
    Long filaments - LD & $1.11 \pm 0.39$ & $-6.09 \pm 0.32$ & $0.31$\\
\hline
 \end{tabular}
\end{table*}

The resulting scaling relations are presented in Fig.~\ref{Fig:Scalings}, and the values of the fitting parameters are reported in Table~\ref{Table:FIT_T} and \ref{Table:FIT_P} for temperature and pressure, respectively.
As for the 1D distributions, the 2D contours in the $\langle \rho_\mathrm{gas} \rangle - \langle T \rangle$ and $\langle \rho_\mathrm{gas} \rangle - \langle P \rangle$ planes clearly reflect the hierarchy between the considered structures: clusters are found in the upper right corner of the plots (where the density, temperature, and pressure values are the highest), and groups, short, and long filaments are distributed in a decreasing diagonal (in this respective order) down to the lower left corner, where the values are the lowest. 
Nevertheless, a closer inspection of the resulting fitting parameters shows that the structures do not follow a perfect diagonal in the $\langle \rho_\mathrm{gas} \rangle - \langle T \rangle$ and $\langle \rho_\mathrm{gas} \rangle - \langle P \rangle$ planes. Indeed, each structure possesses a different value of their slope $\mathrm{a}$. Despite the values being encompassed in a tight range ($\mathrm{a} \in [0.30, 0.70]$ for temperature and $\mathrm{a} \in [1.11, 1.81]$ for pressure, see Tables~\ref{Table:FIT_T} and \ref{Table:FIT_P}), a clear steepening of the slope is seen in structures with decreasing density. We note the large uncertainties in the $\mathrm{a}$ value of the low density long filaments (Long - LD). These are due to the rather broad temperature and pressure ranges exhibited by this sub-set of 112 filaments (see Fig.~\ref{Fig:rho_T_P_distributions}), leading to 2D asymmetrical distributions that are harder to constrain. The differences highlighted above indicate that it is possible to differentiate between the types of cosmic structures by their relation in the $\langle \rho_\mathrm{gas} \rangle - \langle T \rangle$ and $\langle \rho_\mathrm{gas} \rangle - \langle P \rangle$ planes.


\section{\label{Sect:Conclusions}Discussion and conclusions}

In this paper, we analysed the distribution of DM, gas, and stars around the cosmic filaments detected in the TNG300-1 simulation at redshift $z=0$. The different populations of short and long filaments were studied separately, allowing us to probe the effect of the large-scale environments of cosmic filaments (i.e. denser and hotter versus less dense and colder regions) on our results. By computing radial density profiles of the different matter components around filaments, and by deriving baryon fraction profiles of these cosmic structures, we have reached the following conclusions:

\begin{itemize}
    \item We find that DM, gas, and stars are distributed differently around short and long filaments, the former exhibiting always higher density values than the latter (Fig.~\ref{Fig:density_profiles}). DM, gas, and stars show overlapping over-density profiles at all radii except at the filament cores $r < 0.7$ Mpc (Fig.~\ref{Fig:density_FILS_RESCALED}), where the over-density of baryons departs from that of DM. By modelling the over-density profiles with a generalised $\beta$-model, we find that the slopes of the profiles and the radius at which they transition are independent of the large-scale environment (the latter is only reflected in the different density values found in short and long filaments). The slopes and radial scales are rather set by the matter component, hinting at universal processes in filaments, regardless of the population.\\
    
    \item We find that the baryon fraction profiles $f_b$ of short and long filaments present three different regimes (Fig.~\ref{Fig:fraction_b_g_s}): the cores ($r < r_\mathrm{int}$) are baryon depleted with a depletion factor of $Y_\mathrm{b}(r<r_\mathrm{int}) = 0.90^{+0.06}_{-0.06}$ (short) $0.85^{+0.08}_{-0.07}$ (long filaments), at distances of $r \in [r_\mathrm{int}, r_\mathrm{ext}]$ there is an excess (a bump) with respect to the cosmic fraction $\Omega_\mathrm{b} / \Omega_\mathrm{m}$, and 
    at larger distances ($r > r_\mathrm{ext}$) the $f_b$ profiles are flat and consistent with the cosmic value.
    We show that, AGN feedback could contribute to the observed baryon depletion in the cores, given that the kinetic energy injected into the medium by AGNs can be powerful enough to potentially expel gas outside of the gravitational potential wells induced by filaments (Figs.~\ref{Fig:gravitational_potential}, \ref{Fig:E_AGN_profiles} and \ref{Fig:E_RATIO}). 
    We find that the bump in the baryon fraction profiles is a specific signature of the radial accretion of WHIM gas towards filaments (Fig.~\ref{Fig:FG_gas_phases}). Unable to efficiently cool down, this gas assembles at radial distances of $\sim 1$ Mpc from the filament cores. Given that this gas phase is the dominant one in filaments \citep[e.g.][]{GalarragaEspinosa2021}, this bump specific to WHIM is naturally reflected in the total baryon fraction of cosmic filaments.\\
    
    \item We find two main characteristic radii for baryons in cosmic filaments. From the outer to the inner regions, these are $r_\mathrm{ext} \sim 7$ Mpc and
    $r_\mathrm{int} \sim 0.7$ Mpc. Both radii are found to be independent of the filament population, meaning that they are characteristic of all filaments, regardless from their large-scale environment. They are also independent of the resolution of the simulation.
    
    The first radius, $r_\mathrm{ext}$, marks the departure of the baryon fraction from the cosmic value $\Omega_\mathrm{b} / \Omega_\mathrm{m}$. We show that the total baryon fraction enclosed in a cylinder of radius $r_\mathrm{ext}$ yields the cosmic value (Eq.~\ref{Eq:F_rext}). We thus interpret this radius as the radial extent to which baryonic processes taking place in filaments can affect the large-scale distribution of baryons.
    
    The second one, $r_\mathrm{int}$, corresponds to the limit of the gas depletion observed at the cores of filaments. It is defined as the intersection between $f_b$ and $\Omega_\mathrm{b} / \Omega_\mathrm{m}$, and also reflected in the separation of the DM and gas over-density profiles. This radius marks the regime from which the broad diversity of physical mechanisms (e.g. AGN feedback) play a role in shaping the properties of baryons in the cores of cosmic filaments.\\

    \item We find that the distributions of gas density, temperature, and pressure of filaments and clusters clearly reflect the hierarchical structure of the cosmic web (Fig.~\ref{Fig:rho_T_P_distributions}). The densest structures (i.e. clusters) possess higher temperature and pressure values than the least dense ones (i.e. long filaments).
    This is explicitly seen in the scaling relations (Fig.~\ref{Fig:Scalings}), where the different values of the fitting parameters (Tables~\ref{Table:FIT_T} and \ref{Table:FIT_P}) show that the scaling relations in the $\langle \rho_\mathrm{gas} \rangle - T$ and $\langle \rho_\mathrm{gas} \rangle- P$ planes can be used to differentiate between different types of filaments and clusters.\\
\end{itemize}

This work revealed that the relative distribution of baryons (gas + stars) with respect to DM shares the same trends, common radial scales, and similar depletion factors in short and long filaments, i.e. it is independent of the large-scale environments traced by the different populations. Therefore, we show in this work that, contrary to absolute densities, the relative distribution of DM, gas, and stars exhibits a universal behaviour around cosmic filaments, at least at redshift zero.

As a final point, the observed deviations of gas with respect to DM can also be interpreted in terms of gas `bias' in filaments. Previous works such as \cite{Cui2018, Cui2019} have argued that the cosmic skeleton traced by the gas is not biased with respect to the DM one (in the sense that gas follows the same cosmic web as the DM alone), and that the baryonic processes have almost no impact on large-scale structures in terms of their classification in nodes, filaments, walls and voids.
Here, we somewhat nuance this by showing that gas follows DM in an `un-biased' way only down to a certain filament radius, $r_\mathrm{ext}$. Although the bias is small at the intermediate distances of $r \in [r_\mathrm{int}, r_\mathrm{ext}]$ (where the bump lies), it is definitely non-negligible at the cores of filaments, i.e. at distances lower than $r_\mathrm{int}$. This is seen in both the over-densities (Fig.~\ref{Fig:density_FILS_RESCALED}) and gas fraction profiles (Fig.~\ref{Fig:fraction_b_g_s}) presented in this work.\\

All the results of this work have been derived using the IllustrisTNG simulation. Of course, it would be interesting to perform this study using the following: \textit{(i)} higher resolution large-scale hydro-dynamical simulations, which would allow us to probe the small-scale filaments and build a complete picture of the multi-scale filamentary network of the cosmic web, and \textit{(ii)} other simulations with different baryonic models (and non-radiative runs) in order to better assess the relative efficiency of the mechanisms responsible of the distribution of baryons with respect to DM, in cosmic filaments of different environments. These studies, aiming at bridging the gap between the small (kpc) and large (Mpc) scales, can be complemented by an analysis of the velocity field \citep[see e.g.][]{Zhu2017_mass_and_vel_CW}. This will enable a direct insight on the dynamics of matter that is required, for example, to probe the effect of AGNs at large-scales. These studies are the object of future projects.


\begin{acknowledgements}
We would like to thank the anonymous referee for his/her useful comments and suggestions. This research has been supported by the funding for the ByoPiC project from the European Research Council (ERC) under the European Union’s Horizon 2020 research and innovation program grant agreement ERC-2015-AdG 695561. (ByoPiC, \url{https://byopic.eu}). The authors acknowledge the very useful comments and discussions with all the members of the ByoPiC team (\url{https://byopic.eu/team/}). We also thank the IllustrisTNG team for making their data publicly available, and for creating a user friendly and complete website.
\end{acknowledgements}

\bibliography{main} 

\begin{appendix}

\section{\label{Appendix:Subsampling}Impact of the sub-sampling of particles and cells}

\begin{figure}[h!]
    \centering
   \includegraphics[width=0.5\textwidth]{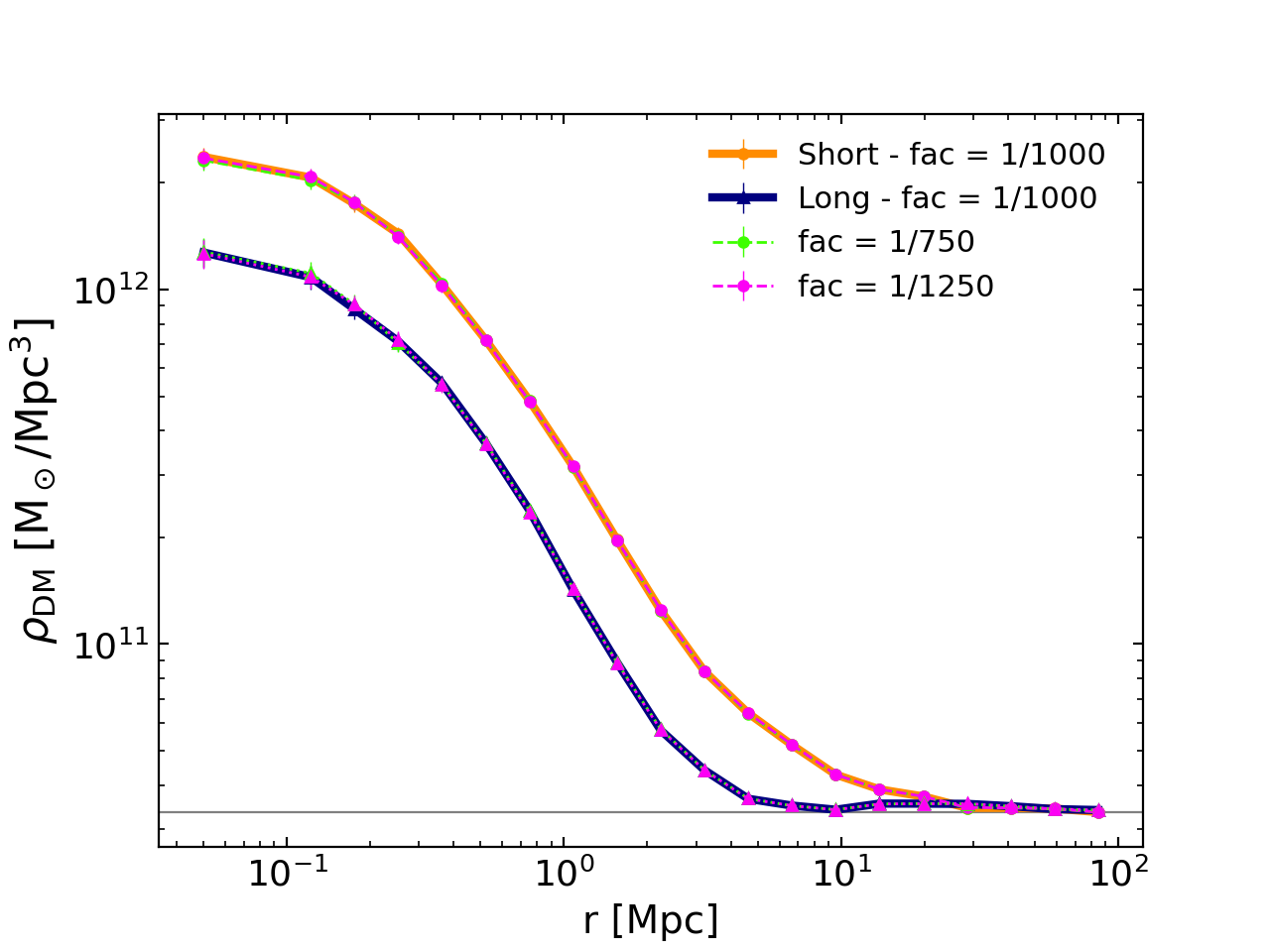}
   \caption{Mean DM density profiles of short and long filaments, for different sub-sampling factors of the DM particle distribution of the TNG300-1 simulation.}
    \label{Fig:APP_pro_SL_DifferentSubsampling}
    \end{figure}

As presented in Sect.~\ref{SubSect:Analysed_data_sets}, in order to perform the analysis of the density profiles within reasonable computing time, the DM, gas, and star data sets of the TNG300-1 simulation are sub-sampled (by randomly taking one out of 1000 particles or cells). 
Fig.~\ref{Fig:APP_pro_SL_DifferentSubsampling} shows the resulting average DM density profiles of short and long filaments for DM datasets having different sub-sampling factors. We can see that the profiles derived from the $1/1000$ dataset are essentially the same as those computed from slightly lower and higher sampling factors (i.e. $1/750$ and $1/1250$). Thus, a reasonable value of sub-sampling of the initial datasets does not affect the resulting densities.\\

\begin{figure}
    \centering
    \includegraphics[width=0.5\textwidth]{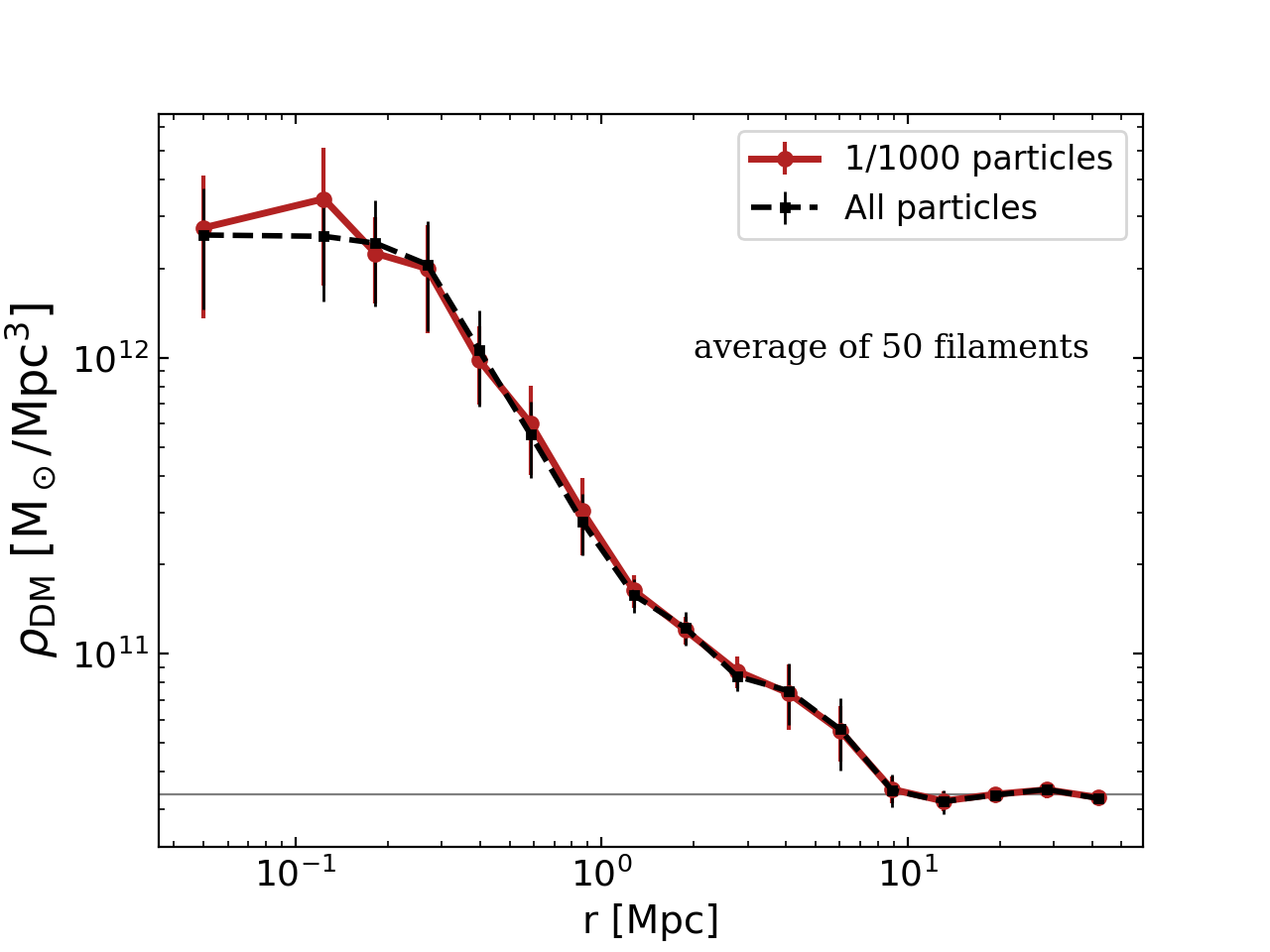}
   \includegraphics[width=0.5\textwidth]{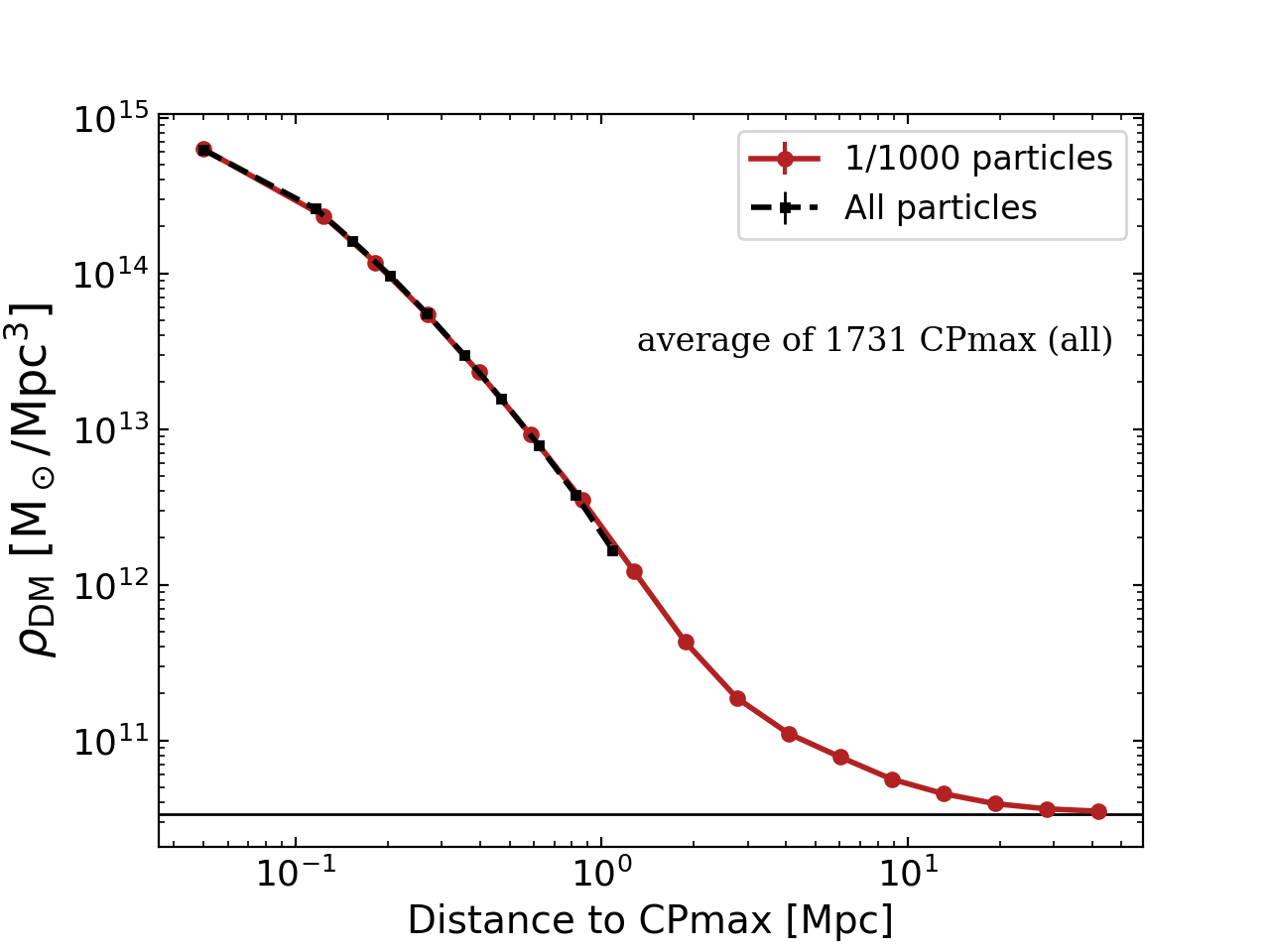}
   \caption{Mean DM density profiles around a set of 50 randomly selected filaments (top panel), and around all the DisPerSE CPmax points of the TNG300-1 simulation (bottom panel). The black dashed profiles correspond to those resulting from the full DM data set, while the red solid curves show the results derived from the fiducial sub-sampled DM data set (where one out of 1000 particles is randomly selected)}.
    \label{Fig:APP_pro_AllDM_vs_downsample}
    \end{figure}

    \FloatBarrier
In order to check that this procedure does not introduce a systematic bias in the density values, we compared the densities estimated from the fiducial sub-sampled data set (factor $1/1000$) to those derived from the full DM distribution of the TNG300-1 simulation (i.e. all the particles around a given filament were considered). Due to limited computing time and memory, this comparison focused only on 50 randomly selected filaments. Their average profiles computed from the full DM distribution and from the sub-sampled data set are presented in the top panel of Fig.~\ref{Fig:APP_pro_AllDM_vs_downsample}.

The lower panel of this figure shows the average profile of the total number (i.e. 1731) of maximum density critical points (CPmax) of the DisPerSE catalogue. In fact, the (assumed) spherical symmetry of the CPmax simplifies the computations of the density profiles with respect to the case of filaments, significantly lowering the memory requirements.

For both filaments and nodes, we see that the DM profiles computed from the full particle distribution (black profiles) are in agreement with those derived from the sub-sampled data set (red). The results in Fig.~\ref{Fig:APP_pro_AllDM_vs_downsample} thus demonstrate that the sub-sampling scheme adopted in this work does not bias the density estimates with respect to the full particle distribution.

    \FloatBarrier

\section{\label{Appendix:Plummer}Results of the fitting of Plummer profile}

This appendix shows the results of the fit of the over-density profiles to the Plummer profile described in Eq.~\ref{Eq:APP_Plummer}:
\begin{equation}\label{Eq:APP_Plummer}
    1 + \delta_i(r) = \frac{1 + \delta^0_i}{ \left(1 + \left(\frac{r}{r_0}\right)^2 \right)^2 }.
\end{equation}

\begin{figure*}[h!]
    \centering
   \includegraphics[width=0.9\textwidth]{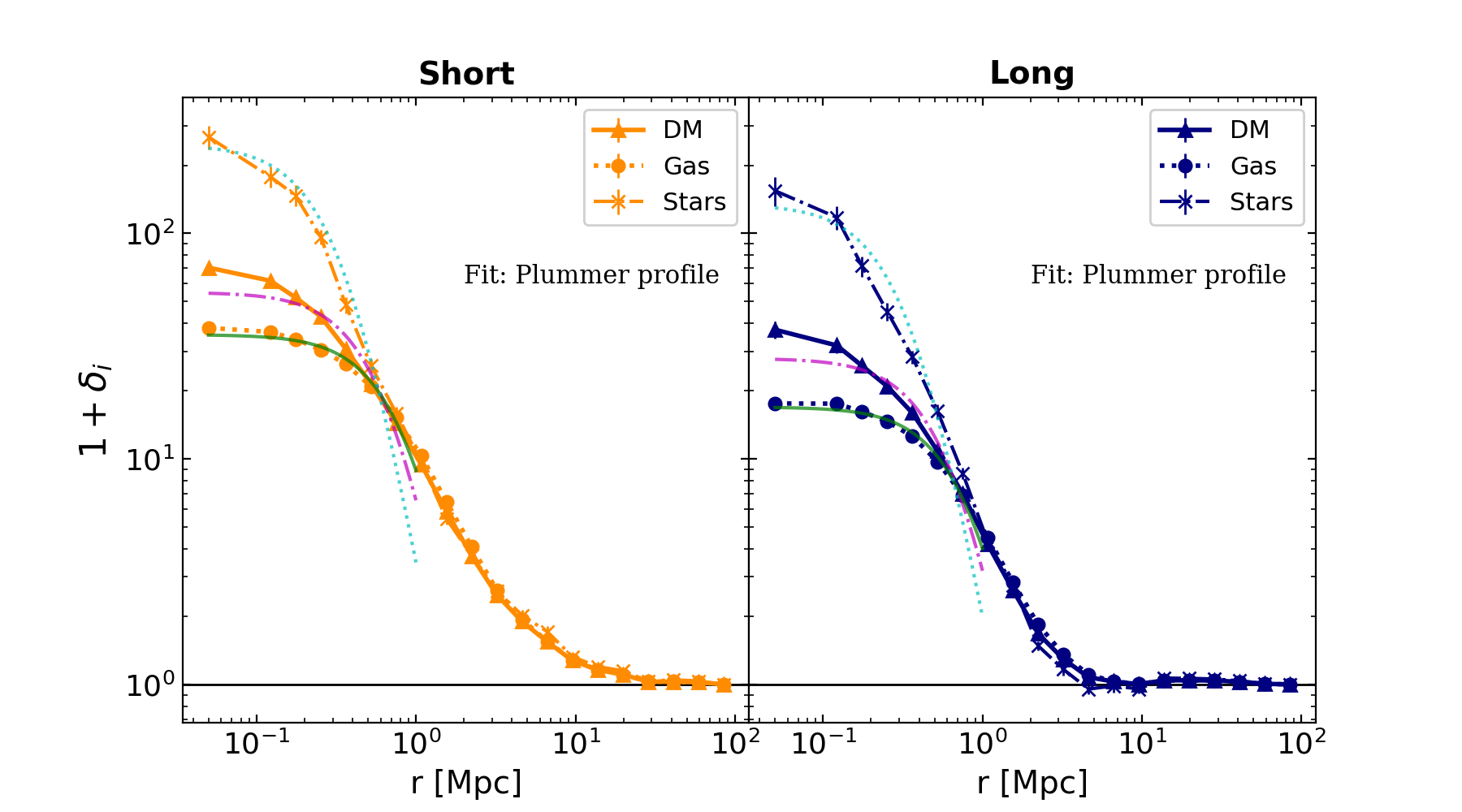}
   \caption{Results of the fitting of the mean over-density profiles of short and long filaments to the Plummer profile of Eq.~\ref{Eq:APP_Plummer}. The fit-curves are plotted using the parameters of Tables \ref{Table:APP_PlummerS} and \ref{Table:APP_PlummerL}.}
    \label{Fig:APP_PlummerFit}
    \end{figure*}

The obtained fit-curves are presented in Fig.~\ref{Fig:APP_PlummerFit} and the resulting parameters for short and long filaments are presented respectively in Tables~\ref{Table:APP_PlummerS} and \ref{Table:APP_PlummerL}. These results are discussed in the main text (see Sect.~\ref{SubSect:Fit_OV}).

    \FloatBarrier

    \begin{table}[h!]
\caption{Fitting results of the over-density profiles of short filaments (left panel of Fig.~\ref{Fig:APP_PlummerFit}) to the Plummer model of Eq.~\ref{Eq:APP_Plummer}.}
\label{Table:APP_PlummerS}     
\centering  
\begin{tabular}{ c | c c | c }
 \hline\hline  
     & $(1 + \delta_0)$ & $r_0$ [Mpc] & $\chi^2_\nu$ \\ \hline
    Dark matter & $54.74 \pm 5.54$ & $0.73 \pm 0.06$ & $11.73$\\
    Gas & $35.55 \pm 1.17$ & $1.00 \pm 0.03$ & $2.52$ \\ 
    Stars & $247.43 \pm 1.79$ & $0.37 \pm 0.05$ & $13.95$ \\
    \hline
 \end{tabular}
\end{table}
\FloatBarrier

\begin{table}[h!]
\caption{Fitting results of the over-density profiles of long filaments (right panel of Fig.~\ref{Fig:APP_PlummerFit}) to the Plummer model of Eq.~\ref{Eq:APP_Plummer}.}
\label{Table:APP_PlummerL}     
\centering  
\begin{tabular}{ c | c c | c }
 \hline\hline  
     & $(1 + \delta_0)$ & $r_0$ [Mpc] & $\chi^2_\nu$ \\ \hline
    Dark matter & $27.90 \pm 2.58$ & $0.71 \pm 0.05$ & $6.10$ \\
    Gas & $16.98 \pm 0.49$ & $0.97 \pm 0.03$ & $0.84$ \\
    Stars & $134.40 \pm 8.52$ & $0.37 \pm 0.05$ & $16.53$ \\
    \hline
 \end{tabular}
\end{table}
\FloatBarrier

\section{\label{Appendix:POTENTIALS}Further details on the derivation of $\phi$ and $\varepsilon_\mathrm{AGN}$}

In this Appendix we show the steps of the derivation of the gravitational potential $\phi$ (Eq.~\ref{Eq:Phi_fils}), and of the energy injected into the medium by AGN feedback by unit of gas mass (Eq.~\ref{Eq:Eagn}).

Firstly, the gravitational potential energy of a test particle of mass $m_\mathrm{part}$ falling towards the filament is:
\begin{equation}
    E_p(R) = m_\mathrm{part} \,\, \phi(R).
\end{equation}
We computed $\phi$ by applying the Gauss theorem to the total matter over-density, $\Delta \rho_\mathrm{TOT} = \rho_\mathrm{TOT} - \rho_\mathrm{bkg}$, in the closed volume defined by a cylinder whose axis corresponds to the filament spine. In the following, let us assume that the matter density is invariant by rotation around the axis of the filament, and by translation along this axis, so that $\Delta \rho_\mathrm{TOT}$ depends only on the radial coordinate.
Within this hypothesis, the result of the Gauss theorem in polar coordinates reads
\begin{equation}\label{Eq:gauss}
    \overrightarrow{g} = \frac{- 2G}{r} \int_0^r \, \mathrm{d}r' \, r' \, \Delta \rho_\mathrm{TOT}(r') \, \overrightarrow{e_r},
\end{equation}
where $\overrightarrow{g}$ is the gravitational field and  $G$ is the universal gravitational constant.
The gravitational field $\overrightarrow{g}$ is just the gradient of the potential $\phi$. Therefore, the expression above allows us to compute $\phi$ by numerically integrating the equation $\overrightarrow{g} = - \overrightarrow{\nabla} \phi$, which yields the result of Eq.~\ref{Eq:Phi_fils}.\\

\begin{figure}[h!]
    \centering
   \includegraphics[width=0.5\textwidth]{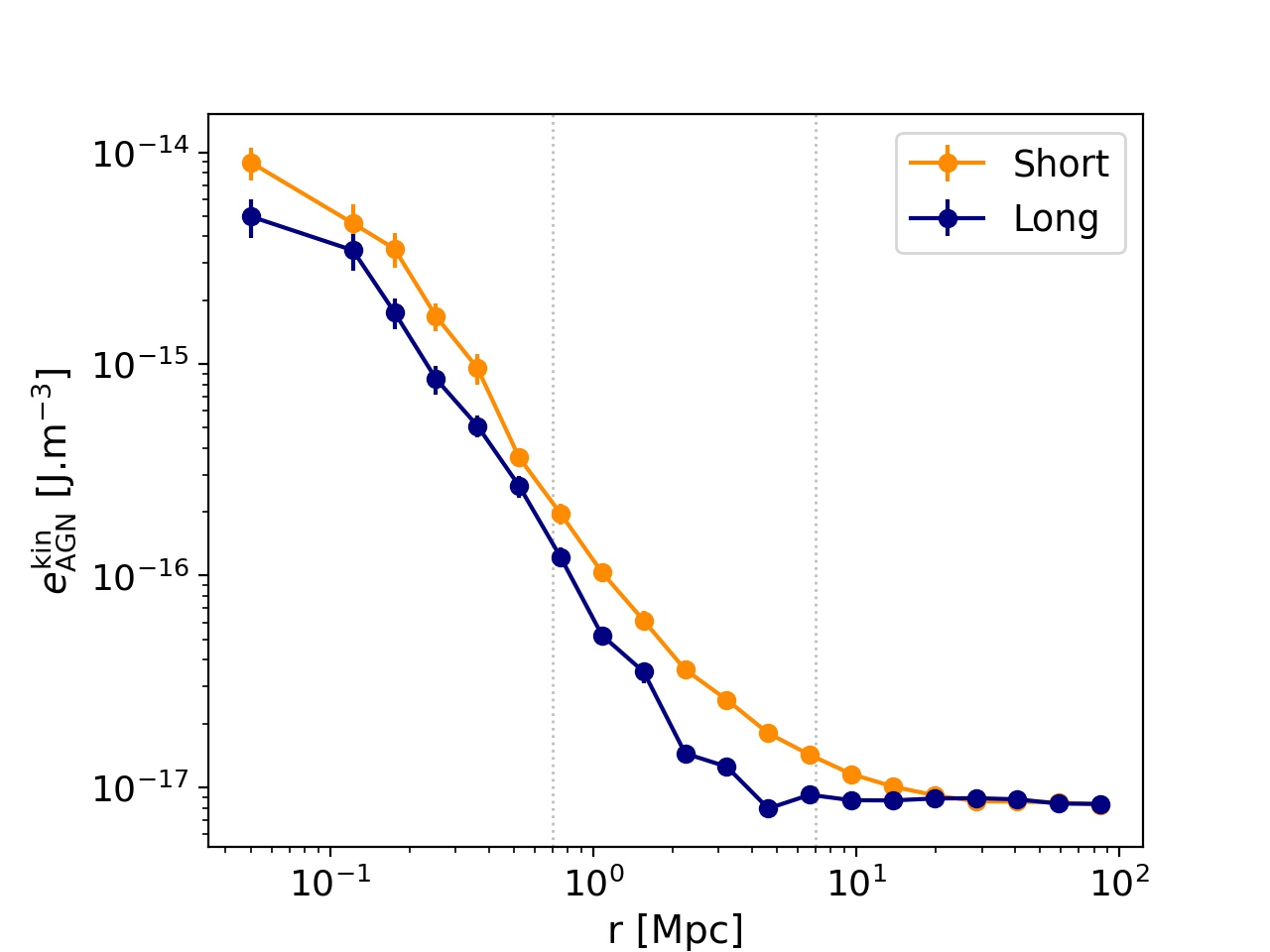}
   \caption{Average radial profiles of kinetic energy per unit volume injected by AGNs into the medium. The inner and outer vertical dashed lines correspond respectively to the $r_\mathrm{int}$ and $r_\mathrm{ext}$ radii.}
    \label{Fig:APP_eAGN_profiles}
    \end{figure}

Secondly, the amount of energy (by unit of gas mass) injected by AGN feedback events into a cylindrical shell of thickness $|r_{k-1} - r_{k}|$ around the axis of the filament is defined as
\begin{equation}\label{Eq:Eagn_def}
    \varepsilon_\mathrm{AGN}(R) \equiv \frac{E_\mathrm{AGN}^\mathrm{inj}(R)}{M_\mathrm{gas}(R)},
\end{equation}
where $R \in [r_{k-1}, r_{k}]$.
Assuming that filaments are cylinders of length $L_\mathrm{fil}$, homogeneous along their axis, the numerator in this equation is given by:
\begin{equation}\label{Eq:Eagn_numerator}
    E_\mathrm{AGN}^\mathrm{inj}(R) = \int_0^{L_\mathrm{fil}} \mathrm{d}l  \int_0^{2\pi} \mathrm{d}\theta \int_{r_{k-1}}^{r_{k}} \mathrm{d}r \,r \,\, e_\mathrm{AGN}^\mathrm{kin}(r),
\end{equation}
where $e_\mathrm{AGN}^\mathrm{kin}$ is the energy per unit volume, presented in Fig.~\ref{Fig:APP_eAGN_profiles}.

The denominator of Eq.~\ref{Eq:Eagn_def} is just the integral of the gas density over the cylindrical volume:
\begin{equation}\label{Eq:Eagn_denominator}
    M_\mathrm{gas}(R) = \int_0^{L_\mathrm{fil}} \mathrm{d}l  \int_0^{2\pi} \mathrm{d}\theta  \int_{r_{k-1}}^{r_{k}} \mathrm{d}r \, r \,\, \rho_\mathrm{gas}(r).
\end{equation}
The ratio between Eq.~\ref{Eq:Eagn_numerator} and Eq.~\ref{Eq:Eagn_denominator} yields the final expression of $\varepsilon_\mathrm{AGN}$ of Eq.~\ref{Eq:Eagn}.

    \FloatBarrier
    
\end{appendix}


\end{document}